\documentclass[sigconf]{acmart}

\AtBeginDocument{%
  \providecommand\BibTeX{{%
    \normalfont B\kern-0.5em{\scshape i\kern-0.25em b}\kern-0.8em\TeX}}}

\usepackage[utf8]{inputenc} 
\usepackage[T1]{fontenc}    
\usepackage{hyperref}       
\usepackage{url}            
\usepackage{booktabs}       
\usepackage{amsfonts}       
\usepackage{nicefrac}       
\usepackage{microtype}      
\usepackage{xcolor}         
\usepackage{graphicx}
\usepackage{anyfontsize}
\usepackage{subcaption}
\usepackage{caption} 
\usepackage{gensymb}
\usepackage{enumitem}

\newcommand{\xref}[1]{\S\ref{#1}}

\newcommand{\squishlist}{\begin{itemize}[itemsep=1pt,parsep=2pt,topsep=3pt,partopsep=0pt,leftmargin=0em, itemindent=1em,labelwidth=1em,labelsep=0.5em]}
\newcommand{\squishend}{\end{itemize}}

\newenvironment{todo-env}{\par\color{red}}{\par}
\newenvironment{help-env}{\par\color{blue}}{\par}
\newenvironment{ready-for-review}{\par\color{violet}}{\par}

\copyrightyear{2024}
\acmYear{2024}
\setcopyright{rightsretained}
\acmConference[CHI '24]{Proceedings of the CHI Conference on Human Factors in Computing Systems}{May 11--16, 2024}{Honolulu, HI, USA}
\acmBooktitle{Proceedings of the CHI Conference on Human Factors in Computing Systems (CHI '24), May 11--16, 2024, Honolulu, HI, USA}
\acmDOI{10.1145/3613904.3642057}
\acmISBN{979-8-4007-0330-0/24/05}

\title{Look Once to Hear: Target Speech Hearing    with Noisy Examples}

 \author{Bandhav Veluri}\authornote{Co-primary student authors}
  \affiliation{Paul G. Allen School, University of Washington, Seattle, WA  
 \country{USA}
 }
 \email{bandhav@cs.washington.edu}

 \author{Malek Itani}\authornotemark[1]
  \affiliation{Paul G. Allen School, University of Washington, Seattle, WA  
 \country{USA}
 }
 \email{malek@cs.washington.edu}

 \author{Tuochao Chen}
 \affiliation{Paul G. Allen School, University of Washington, Seattle, WA  
 \country{USA}
 }
 \email{tuochao@cs.washington.edu }

 \author{Takuya Yoshioka}
 \affiliation{AssemblyAI, \\ San Francisco, CA
  \country{USA}
 }
 \email{takuya.yoshioka@ieee.org}

 \author{Shyamnath Gollakota}
 \affiliation{Paul G. Allen School, University of Washington, Seattle, WA
   \country{USA}
 }
\email{gshyam@cs.washington.edu}



\begin{document}


\begin{abstract}

In crowded settings, the human brain can focus on speech from a target speaker, given prior knowledge of  how they sound. We introduce a novel intelligent hearable system that achieves this capability, enabling target speech hearing to ignore  all interfering speech and noise, but the target speaker. A na\"ive approach is to  require a clean speech example  to enroll the target speaker. This is however not well aligned with the hearable application domain since  obtaining a clean  example is challenging in  real world scenarios, creating a unique user interface problem.  We present the first enrollment interface where the wearer looks at the target speaker for a few seconds to  capture a single, short, highly noisy, binaural example  of the target speaker. This noisy example is used for enrollment and subsequent speech extraction in the presence of interfering speakers and noise.
Our  system achieves a  signal quality improvement of 7.01~dB using less than 5 seconds of noisy enrollment audio and can process 8~ms of audio chunks  in 6.24~ms on an embedded CPU. Our user studies demonstrate generalization to real-world static and mobile speakers in previously unseen indoor and outdoor multipath environments. Finally, our  enrollment interface for noisy examples does not cause performance degradation compared to clean examples, while being convenient and user-friendly. Taking a step back, this paper takes an important step towards  enhancing the human auditory perception with artificial intelligence. We provide code and data at: {\textcolor{blue}{\url{https://github.com/vb000/LookOnceToHear}}}.
    
\end{abstract}


\begin{teaserfigure}
\centering
  \includegraphics[width=0.8\textwidth]{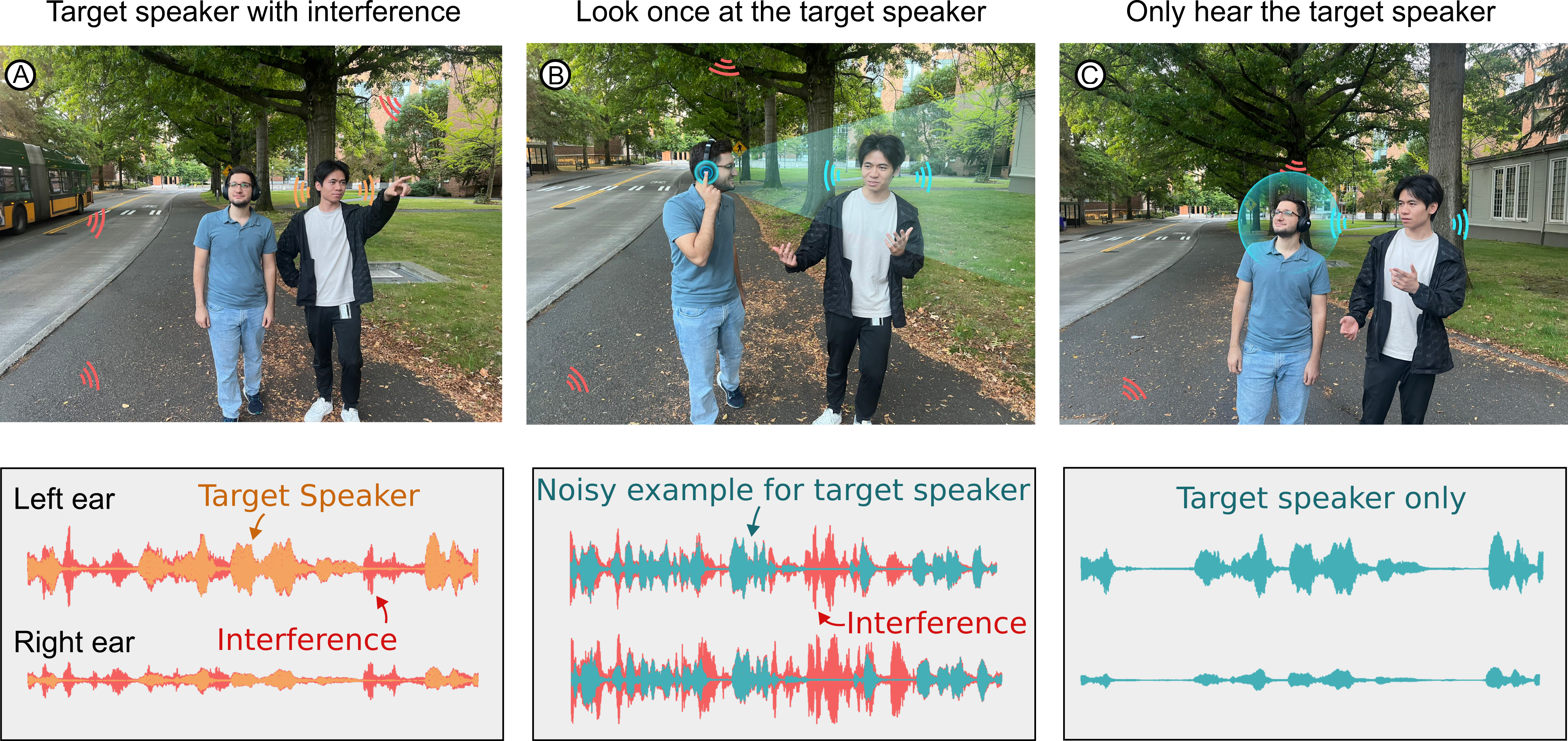}
  \caption{"Look once to hear" is an intelligent   hearable system where  users  choose
to hear a target speaker by  looking at them  for a few seconds. (A) Two users are walking near a noisy street,  (B) the wearer looks at the target speaker for a few seconds to capture a  noisy binaural audio example, which is used to learn the speech traits of the target speaker,  and (C) the hearable extracts the target  speaker and removes interference, even when the wearer is no longer looking at the target speaker. }
  \label{fig:teaser}
\end{teaserfigure}

\maketitle

\section{Introduction}



The past decade has witnessed two key technological trends. First, there have been significant advances in noise-canceling headsets and earbuds capable of better suppressing all environmental  sounds \cite{airpods,history,review}. Second, deep learning is enabling promising human-like intelligence across various domains~\cite{bubeck2023sparks,nature11}. These two trends present  opportunities for creating    the future of intelligent hearables, with real-world capabilities that so far have been in the realm of science fiction. In this paper, we explore a novel capability for hearables --- {\it target speech hearing} ---  that allows users to choose to hear target speakers based on user-selected target speaker characteristics, such as speech traits.

Specifically, we explore the following question: can we look at a target speaker within a crowd just once, extract their unique speech traits, and subsequently employ these traits to exclusively listen to that speaker, while filtering out other voices and background noise?
 A positive answer could enable novel hearable applications that are currently not possible. For example, imagine a scenario in which a user seeks to hear only the tour guide's narration during a guided tour amidst the surrounding chatter and ambient noise while enjoying the tour sights. Alternatively, picture a leisurely stroll with a colleague along a cacophonous street, wanting to hear only their conversation and block out other sounds. Or think about being on a crowded bus, desiring to hear your friend talk while simultaneously gazing out of the window. While today's noise-canceling headphones have seen significant improvements in canceling out {\it all} sounds, they cannot selectively pick speakers based on their speech traits. These use cases, however, require not only using noise-canceling headsets to remove all sounds but also playing  only the target speech back into the hearables.


The latter, which we call  target speech hearing, is a new  capability  for general-purpose  hearable devices. Existing deep learning approaches for the problem of target speech extraction require  prior clean audio examples of the target speaker~\cite{tse-overview}. These clean examples are utilized by a neural network to learn the characteristics of the target speaker, which are subsequently employed to separate their speech from that of other concurrent speakers. The challenge lies in the fact that this problem formulation does not align well with our target  hearable application domain. Specifically, in all the previously described  use cases, obtaining a clean example  signal of the target speaker (e.g., tour guide) is difficult since the target speaker may always be in a noisy environment, with interference from other speakers.

Providing clean target speaker examples for enrollment is essentially a user interface problem, and hence requires the design of an  intelligent hearable system that takes into account the constraints of a user-friendly interface. In this paper, we introduce   the concept of target speech hearing on hearable devices with noisy examples. To achieve this, rather than expecting users to collect input examples of the target speaker in a noise-free environment in the absence of  any other speakers, we show for the first time how one can enable target speech extraction using noisy binaural enrollments in the presence of other concurrent interfering  speakers.


As shown in Fig.~\ref{fig:teaser}, the wearer looks at the target speaker for a few seconds and captures  binaural audio,  using two microphones, one at each ear. Since during this short enrollment phase, the wearer is looking in the direction of the target, the signal corresponding to the target speaker is aligned across the two binaural microphones, while the other interfering speakers are likely to be in a different direction and are therefore not aligned. We employ a neural network to learn the characteristics of the target speaker using this sample-aligned binaural signal and separate it from the interfering speaker using direction information. Once we have learnt the characteristics of the target speaker (i.e., target speaker embedding vector) using these noisy binaural enrollments, we subsequently input the  embedding vector into a different neural network to extract the target speech from a cacophony of speakers. The advantage of our approach is that the wearer only needs to look at the target speaker for a few seconds during which we enroll the target speaker. Subsequently, the wearer can look in any direction, move their head, or walk around while still hearing  the target speaker.\footnote{In contrast,   directional hearing~\cite{directional}  focuses on speech from a specific direction. However, this approach is not well-suited to our application scenarios, as users do not continuously look at the target speaker, the target speaker may have long pauses in their speech making  continuous direction tracking challenging, and the direction  can change as they or the user move their head to look elsewhere (e.g., tour sights).} 


To make this idea practical, we make multiple     contributions: 

\squishlist

\item {\bf Enrollment networks with noisy examples.} We design and compare two different enrollment networks --- a beamformer network and a knowledge distillation network  (see~\xref{sec:enrollment}) ---  to effectively  generate a speaker embedding vector that captures the traits of the target speaker using the short binaural  noisy example.
\item {\bf Real-time embedded target speech hearing network.} We use the generated embedding to subsequently extract the target's speech using an optimized  network that runs in real-time on an embedded IoT CPU. To do this, we start with the state-of-the-art speech separation network, TFGridNet~\cite{tfgridnet}, which cannot run in real-time on our embedded device. We introduce various model and system-level optimizations in~\xref{sec:realtime} to achieve a light-weight target speech hearing network that  runs in real-time on  embedded CPUs.
\item {\bf Generalization to real-world multipath, HRTF and mobility.}  We present a training methodology that uses only synthetic data and yet allows our system to generalize to real-world unseen target and interfering speakers and their  head-related transfer functions (HRTFs). Further, we explicitly train  with multipath to generalize to both indoor and outdoor environments. We also introduce a fine-tuning mechanism that addresses  moving sources and sudden changes in the listener’s head orientation (upto 90$^{\circ}$/s angular velocity). This also  allows the system to handle up to 18$^{\circ}$  error in the listener's head orientation during  enrollment  (see~\xref{sec:training}).
\squishend

We build an end-to-end hardware system that integrates a noise-canceling headset (Sony WH-1000XM4), a pair of binaural microphones (Sonic Presence SP15C) with our real-time target speech hearing network running on an embedded IoT CPU (Orange Pi 5B). The embedded device reads audio chunks from the microphones, which we process on-device and play back to the headset. Our average model inference time was 6.2~ms to process 8~ms audio chunks, making it a real-time system with a total end-to-end latency of 18.24 ms. Our results are as follows.

\squishlist
\item  Compared to clean example enrollments, the beamformer  network for   noisy example enrollments resulted in 2.9 dB performance drop. In contrast, the knowledge distillation network resulted in only a 0.4 dB drop in performance compared to  clean examples  (see~\xref{sec:benchmark}), while using only 1-4 second noisy enrollments.
\item Our system  generalized to 9 real-world {settings} that span  different motion scenarios, indoor and outdoor environments as well as different wearer postures with 8 participants using our hardware. Our design does not  require any training data
collection with our hearable hardware.
\item In a user study with 21 participants who spent over 420 minutes rating the target-speaker output by our hardware system from real-world  indoor and outdoor environments, our system achieved a higher mean opinion score and
interference removal for the target speaker than the raw unprocessed input.
\item Across nine  participants who compared three  interfaces for noisy enrollments ---  push button on headphone, touchpad on headphone, and virtual button on a smartphone --- participants expressed preference for the push button because of its good haptic feedback.

\squishend

{Imbued with embedded intelligence, our work envisions  hearables that  allow wearers  to manipulate their acoustic surroundings in real-time to customize their auditory experience based on user-defined characteristics like speech traits. By open sourcing the  code and datasets,  our work may help further future research among HCI and machine learning researchers on designing algorithms and systems around target speech hearing.}





\section{Background and Related Work}

To the best of our knowledge,  prior work has not explored the concept of target speech hearing with noisy enrollments. Below we describe related work in acoustic machine learning, systems and interaction mechanisms for hearable devices.

 {\bf AI-based hearable systems.} Recent hearable systems such as Clearbuds~\cite{chatterjee2022clearbuds} can enhance the speech of  the user wearing the earbuds for telephony applications,  but cannot pick and choose any other target speaker. Further, telephony applications have a delay constraint of 100-200~ms, which is an order of magnitude less constrained than our system.  Our work is also related to  semantic hearing~\cite{semantichearing}. The goal of this work is to pick and choose which classes of sounds the user wants to hear (e.g., car honks, nature sounds). Our work differs from this work in two key aspects. First, all speech is just one sound class in the semantic hearing system. It is not designed to separate different speakers given their speech characteristics. Second, all the sound classes are pre-defined for the semantic hearing system. In contrast,  our proposed target speech hearing system is designed to focus on any target speaker who is not in the training data. To achieve this, we need an enrollment interface to capture a recording of the target speaker so the system during inference knows their speech characteristics.  Finally, there has been recent interest in using EEG signals~\cite{eeg2}, which potentially could be used to identify the target speaker. While prior work has shown some promise using a head EEG scalp with a large number of electrodes, practical designs need custom in-ear EEG hardware~\cite{berkeleyearable} with electrodes placed only inside the ear
 { or unobtrusive multi-channel EEG acquisition from around the ear (cEEGrids~\cite{cEEGrid}).  The effectiveness of identifying the target speaker using in-air or  around-ear dry electrodes~\cite{separationeeg}  in noisy uncontrollable environments requires further investigation and improvements.} In contrast, our design works with existing binaural hearable hardware architectures, and requires only two microphones typical to today's  hearables.

{\bf Noise-canceling  hearables.}  Active noise-canceling headsets and earbuds, including the lightweight in-ear systems like AirPods Pro, can now cancel upto 20-30~dB signal across the audible frequency range~\cite{review,cancel}. These earable systems  cancel all signals by transmitting an anti-noise signal and have much stronger delay requirements than our target speech hearing system. Further, they cannot pick and choose specific speakers based on their  speech characteristics.  Our design leverages these advances in noise-canceling hearables to reduce the amplitude of all external sounds and noise and then introduce back the target speaker through the headset  with end-to-end delays of less than 20~ms. 

{\bf Neural networks for target speech extraction.} The goal here is to extract the speech signal of a target speaker, from a mixture of several speakers, given additional clues to identify the target speaker~\cite{tse-overview}. Prior work has explored three kinds of clues: audio clues from pre-recordings of the target speaker~\cite{speakerbeam,tse-multi1,tse1,tse2,tse4,tse5,tse6}, visual clues using a video recording~\cite{video1} and spatial clues by providing the direction and/or location of the target speaker. Deep learning  has been used for  target speech extraction using only a few seconds of pre-recorded audio of the target speaker~\cite{giri2021personalized,tse11}.  However, all existing target speech extraction approaches, including those that use multiple microphones~\cite{tse-multi1,tse-multi2,tse-multi3,speakerbeam}, require a clean audio recording of the target speaker without any interference from other speakers. In contrast, we introduce the first target speech extraction system that uses noisy enrollments from binaural hearables and addresses this fundamental interface problem of inputting a clean target speaker recording.

Prior work also proposes visual clues for this task~\cite{video1,video2,video3,video4,video5,video6,video7,video8,video10}. However almost none of the existing commercial hearable systems like headsets and earbuds have cameras.  Further, the lack of adoption of head-worn camera systems like Google glasses, might point to a cultural hesitance to such systems~\cite{social}.  Target speech extraction is also related to the more general blind source separation problem~\cite{tfgridnet} where the task is to separate all speakers in a mixture. This is challenging with an unknown number of speakers and with permutations between mapping the model output to the corresponding speakers~\cite{tse-overview}.

\begin{figure*}[t!]
    \centering
    \begin{subfigure}[t]{0.3\textwidth}
        \centering
        \includegraphics[height=2.9in]{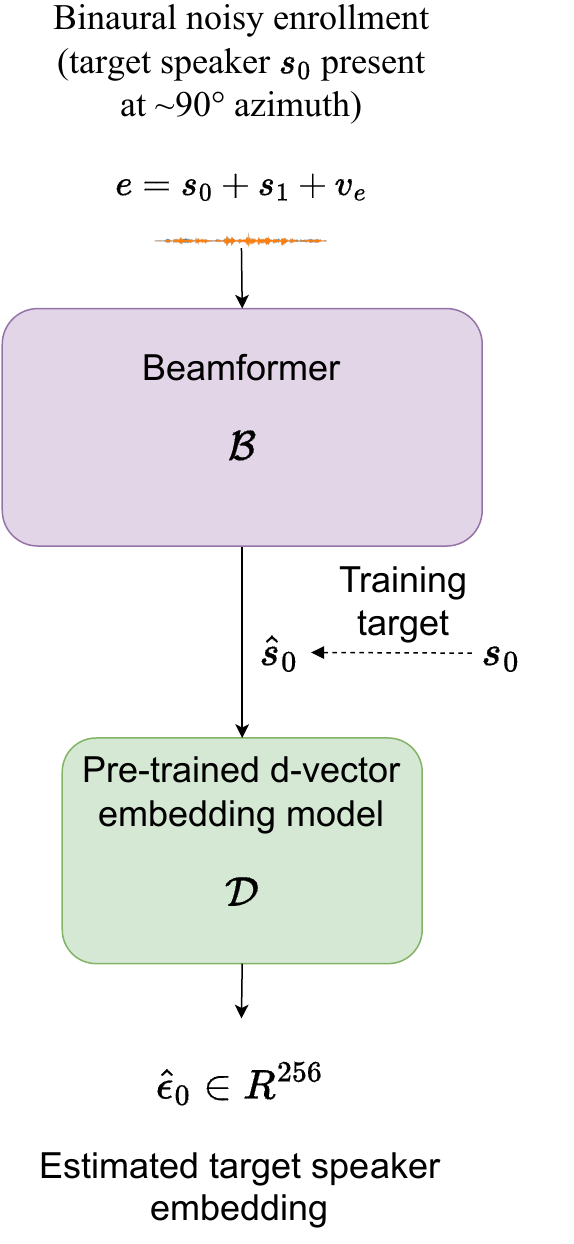}
        \vskip -0.05in
        \caption{}
        \label{beamformer}
    \end{subfigure}%
    ~
    \begin{subfigure}[t]{0.4\textwidth}
        \centering
        \includegraphics[height=2.8in]{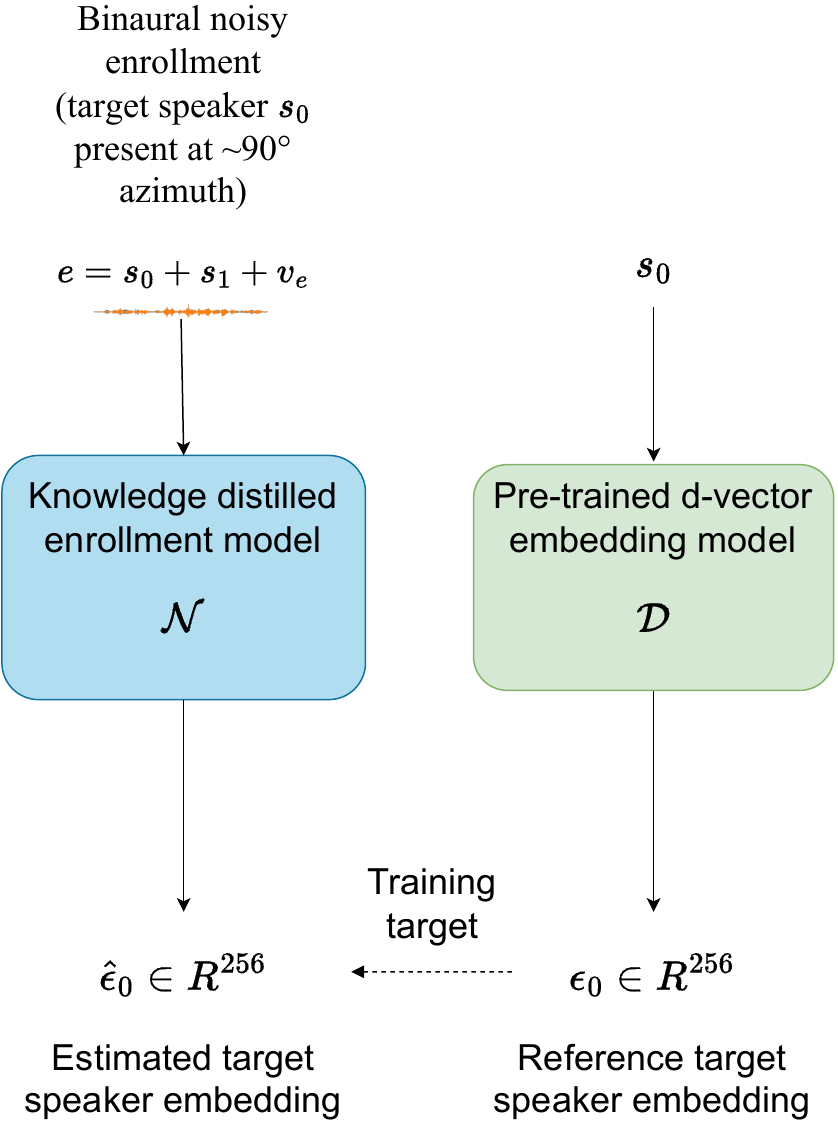}
                \vskip -0.05in
        \caption{}
        \label{fig_kd}
    \end{subfigure}
    ~
    \begin{subfigure}[t]{0.3\textwidth}
        \centering
        \includegraphics[height=3.1in]{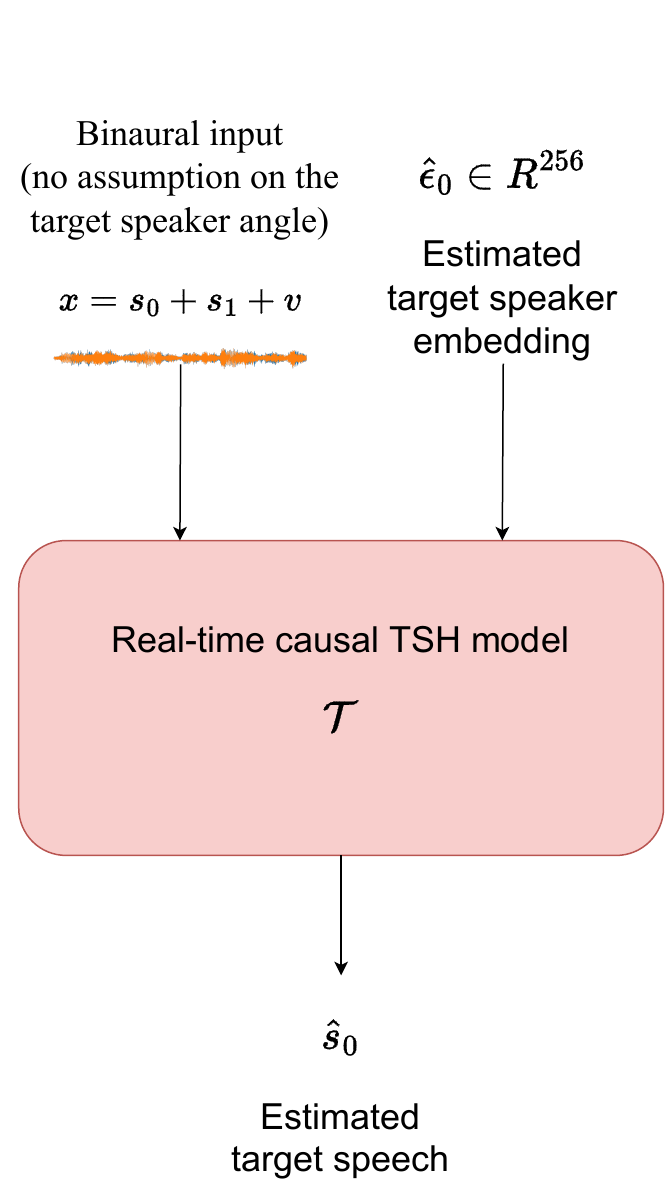}
                \vskip -0.05in
        \caption{}
        \label{tsh}
    \end{subfigure}
            \vskip -0.15in    \caption{\textbf{Target speech hearing with noisy enrollments.} In (a) and (b), we propose two approaches for performing noisy enrollment, assuming the target speaker's azimuthal angle is approximately equal to $90^\circ$. (a) shows the beamformer-based approach, where a beamformer is trained to estimate the target speech signal from the noisy enrollment. The estimated target speech signal is then used to estimate the target speaker's embedding. (b) shows the knowledge-distillation approach, where an enrollment model is trained to estimate the reference d-vector embedding of the target speaker present at $\sim90^\circ$ azimuth. In (c) we use the speaker embedding estimated with one of the two approaches to perform target speech hearing in real-time.}
\label{sys_analysis}
\end{figure*}

{{\bf Beamforming and directional hearing.} Providing the direction  of the target speaker as input is the task of directional hearing or beamforming~\cite{directional,direction1}.  This is performed by jointly processing audio signals captured from multiple microphones to  amplify speech from a specific direction. The traditional approach has been to use statistical signal processing methods~\cite{krim1996two, brayda2015spatially}, which are   computationally light-weight.  Recent work has shown that  neural networks achieve improved performance over signal processing~\cite{souden2010study,jenrungrot2020cone} and can run on-device~\cite{directional}. While we use neural beamformers for the enrollment phase, as described earlier, directional hearing is not well-suited for our  target applications, as users (e.g., in the touring scenario) do not continuously look at the target speaker. Further since speech can have long pauses and is interleaved with other speakers, it is challenging to track the direction of a mobile  target speaker if the user is not continuously looking at them. }

{{\bf Causal, non-causal and real-time models.}  Speech to speech neural networks have been proposed in the context of speech enhancement \cite{dubey2023icassp, DBLP:journals/corr/ParkL16, schroeter2022deepfilternet} and speech separation \cite{luo2019conv, luo2020dual, https://doi.org/10.48550/arxiv.1711.00541}. Most of these models \cite{ subakan2022resource, tfgridnet, hu2020dccrn} support  offline processing of full-length utterances, referred to as non-causal models. There has been recent research on online processing where the model only has access to past information, referred to as causal models. Real-time implementations of such models \cite{directional,han2020realtime, sunohara2017low, giri2021personalized, waveformer} are able to process a second of speech within a second. We  compare our design with causal and real-time  implementations of DCCRN~\cite{hu2020dccrn} and  Waveformer~\cite{waveformer}, which have been used for target sound/speech extraction tasks. }


\section{Target speech hearing with noisy examples}\label{sec:system}

Our key observation is that for hearable applications of deep learning-based target speech extraction \cite{speakerbeam, video10, tse-overview}, it is often impractical to obtain a clean speech sample of the target speaker. In this work, we propose a \emph{target speech hearing (TSH)} system suitable for binaural hearables applications that provides an interface for noisy in-the-wild speech samples, which we refer to as \emph{noisy enrollments}. A noisy enrollment of a speaker of interest would contain two kinds of noise: uncorrelated background noise, and interfering speech. While the background noise can be suppressed with existing methods \cite{hu2020dccrn, icaasp_dns_results}, it is challenging to disambiguate and suppress interfering speech without suppressing the target speech itself, especially when the number of speakers in the scene could be arbitrary. More fundamentally, in a mixture of multiple speakers, it is challenging to know which of them is the intended target speaker.

Our system achieves this disambiguation by leveraging the beamforming capability of binaural hearables. Assuming that the listener would be looking at the target speaker at least for a few seconds, we propose that the listener could use this phase to \emph{enroll} the speaker they want to focus on by letting the hearable know through on-device haptic control or a button click on the phone application. During this phase, since the direct path of the target speaker is equidistant from both ears of the binaural hearable, the application could disambiguate between target and inferring speakers to obtain a representation of the target speaker.

Let $e(t') \in R^2$ be the input binaural signal received by the binaural hearable during the enrollment phase, and $x(t) \in R^2$ be the input binaural signal received during TSH phase, where $t'$ and $t$ corresponds to the time during the enrollment phase and TSH phase, respectively. Then these signals could be decomposed into their component signals:
$$ e(t') = s_0(t') + \Sigma_{k=1}^m s_{ek}(t') + v_e(t')$$
$$ x(t) = s_0(t) + \Sigma_{k=1}^n s_{k}(t) + v(t) $$

Here, $s_0 \in R^2$ corresponds to the target speaker, $s_{e1},...,s_{em} \in R^2$ correspond to interfering speakers during the enrollment phase, and $s_{1},...,s_{n}$ correspond to interfering speakers during the TSH phase. Note that the interfering speakers can be the same or different during the two phases.  $v_e(t')$ and $v(t)$ represent background noises in the respective phases. Additionally, let $\theta_0$ represent the azimuthal angle of the target speaker, relative to the listener. During the enrollment phase, to achieve disambiguation of the target speaker in the noisy enrollment signal, since the user looks in the direction of the target speaker, we can assume that: $\theta_0(t') \sim \frac{\pi}{2}$, where the x-axis is assumed to pass from the listener's left  to right ear with the midpoint  as the origin. We then formulate the TSH problem as a two-step process:
$$ \hat{\epsilon}_0 = \mathcal{N}(e(t') | \theta_0(t') \sim \frac{\pi}{2})$$
$$ \hat{s}_0(t) = \mathcal{T}(x(t), \hat{\epsilon}_0) $$

Here, $\hat{\epsilon}_0$ corresponds to the target speaker representation computed from the noisy enrollment signal $e(t')$, $\mathcal{N}$ is the neural network estimating the target speaker's representation and $\mathcal{T}$ is the real-time causal target speech hearing network that can run on an embedded  device. In the following subsections, we explain in detail, different architectures we explored for both the enrollment phase and TSH phase.

\subsection{Enrollment interface network}\label{sec:enrollment}

The quality of the target speech extracted by the target speech hearing network, $\mathcal{T}$, has a critical dependence on the discriminative quality of the speaker representation, $\epsilon_0$, provided to it. 
In order to robustly handle various speech characteristics, we leverage the speaker representations computed by large-scale pre-trained models such as \cite{wan2020generalized, koluguri2021titanet}. In this work, we use the open-source implementation of \cite{wan2020generalized} in the  Resemblyzer project \cite{Resemble-Ai}. Given a clean speech utterance of a speaker $s_i(t')$, \cite{Resemble-Ai} uses a long short-term memory (LSTM) network, $\mathcal{D}$, to map the utterance to a unit length 256-dimensional vector $\mathcal{D}(s_i(t')) = \epsilon_i$, where $\epsilon_i \in R^{256}$ and $||\epsilon_i||_2=1$, referred to as a \emph{d-vector embedding}. During the training phase, the LSTM model computes d-vectors optimized such that embedding corresponding to an utterance of a speaker is closest to the centroid of embeddings of all other utterances of the same speaker. This is done while simultaneously maximizing the distance from the centroids of all other speakers in the large-scale speech database used as the training set. In this work, we use d-vector embeddings as reference speaker representations that the noisy enrollment network $\mathcal{N}$ should predict using two approaches.

\textbf{Noisy enrollment with beamforming.} Following the notation in \xref{sec:system}, we note that the d-vector embedding of the target speaker can be obtained with its clean speech example as $\epsilon_0 = \mathcal{D}(s_0(t'))$. If we could estimate the clean speech of the target speaker, provided that the target speaker is present at the azimuthal angle $\theta_0 \sim \frac{\pi}{2}$, we could estimate the d-vector embedding corresponding to the target speaker. Essentially, this is equivalent to beamforming with direction input steered towards the azimuthal angle equal to $\frac{\pi}{2}$. In this work, we follow the \emph{delay and process} approach proposed in several beamforming  works \cite{jenrungrot2020cone, chatterjee2022clearbuds, directional}, where given a target direction and a reference microphone, inputs from other microphones are delayed according to the time it takes for the direct path from the given direction to reach them relative to the reference microphone. In this case, since we assume the direct path is equidistant from both left and right microphones, processing the raw inputs is sufficient to obtain the target speaker. Assuming the beamforming  network is represented as $\mathcal{B}$, the process of noisy enrollment with beamforming could be written as:
$$ \hat{s}_0(t') = \mathcal{B}(e(t') | \theta_0 \sim \frac{\pi}{2})$$
$$ \hat{\epsilon}_0 = \mathcal{D}(\hat{s}_0(t'))$$


In this work we use the state-of-the-art speech separation architecture TFGridNet \cite{tfgridnet} as our beamforming architecture $\mathcal{B}$. Since  enrollment is a one-time operation that does not need to be performed on-device,  we could use the original non-causal implementation of the TFGridNet \cite{tfgridnet} available in the ESPNet\cite{espnet} framework. Following the notation in \cite{tfgridnet}, we used the configuration: $D=64$, $B=3$, $H=64$, $I=4$, $J=1$, $L=4$ and $E=8$ with short-time fourier transform (STFT)  window size set to 128 and hop size set to 64.

\begin{figure*}[t!]
    \centering
    \includegraphics[width=0.99\linewidth]{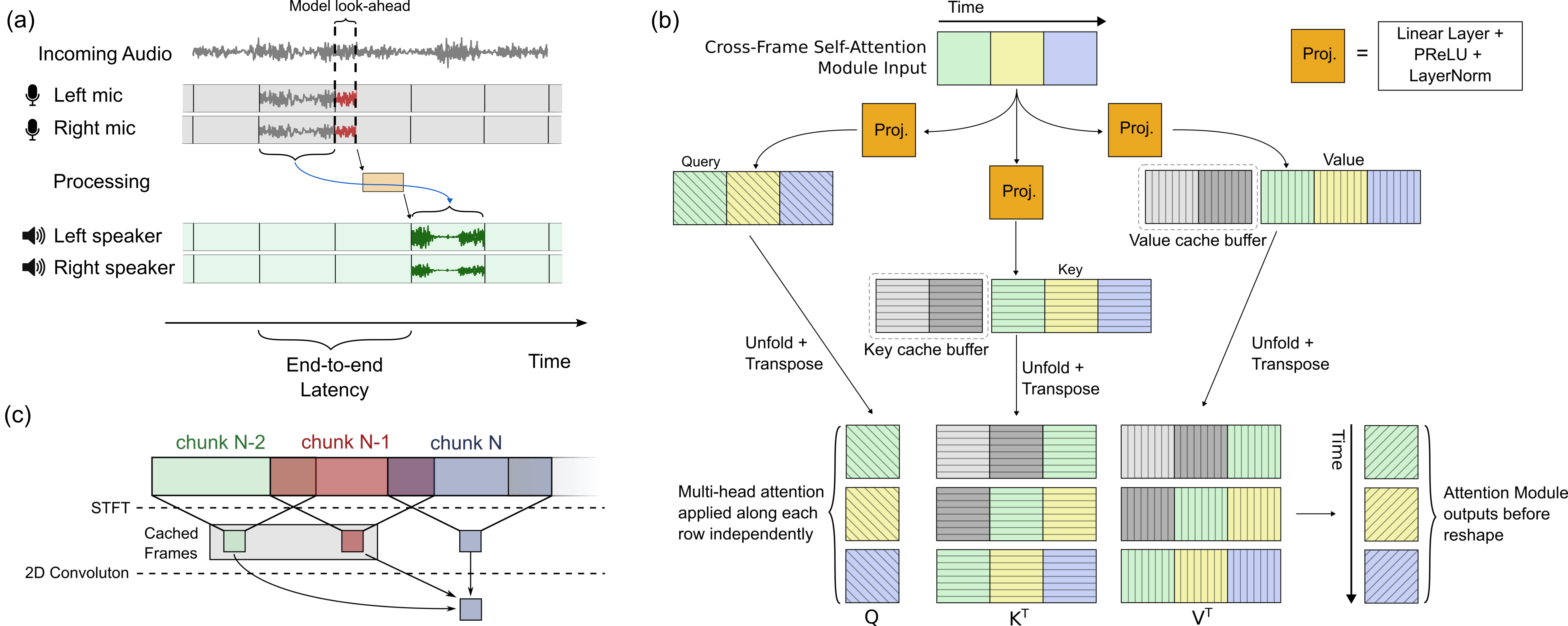}
    \vskip -0.15in
    \caption{{\bf Real-time target speech hearing system. } (a) Decomposition of the system end-to-end latency into various components, including chunk buffering time, model look-ahead time and processing time. (b) Our implementation of a causal cross-frame self-attention module. The figure shows an example procedure that attends to at most 2 frames into the past. The key, query and value tensors are computed from the output of the previous layers, and we concatenate cached values from computations on previous time chunks to quickly compute the updated key and value tensors for multi-head attention. (c) An example where we use cached frames to avoid recomputing the STFT frames for prior chunks when computing the output of
    the 2D convolutional layer.}
    \label{fig:realtime}
\end{figure*}

\textbf{Noisy enrollment with knowledge distillation.} Conversely, we could consider this problem as the noisy enrollment network, $\mathcal{N}$, directly computing the estimated d-vector embedding of the target speaker, $\hat{\epsilon}_0$, given the noisy speech. This would however require us to use a  resource-intensive training process like the one proposed in \cite{wan2020generalized}. To do this efficiently, we train the enrollment network $\mathcal{N}$, using knowledge distillation \cite{hinton2015distilling, 7953145}, where the original d-vector model, $\mathcal{D}$, provides d-vector embeddings computed on clean target speech as ground-truth references. We note that during the training phase, we have access to clean target enrollment speech $s_0(t')$, but we do not assume this during inference. Here, we train the noisy enrollment network $\mathcal{N}$ to minimize the loss function $\mathcal{L}(\hat{\epsilon}_0,{\epsilon}_0)$: 
$$ \hat{\epsilon}_0 = \mathcal{N}(e(t')  | \theta_0 \sim \frac{\pi}{2})$$
$$ \epsilon_0 = \mathcal{D}(s_0(t')) $$
$$ \mathcal{L}(\hat{\epsilon}_0,{\epsilon}_0) = 1 - cos(\angle(\hat{\epsilon}_0, \epsilon_0)) = 1 - \frac{\hat{\epsilon}_0 \cdot \epsilon_0}{||\hat{\epsilon}_0||_2 ||\epsilon_0||_2}$$

To make both our noisy enrollment approaches comparable, we use TFGridNet \cite{tfgridnet} with the same configuration as above, as the noisy enrollment network $\mathcal{N}$, in this approach as well. We modify the architecture to output 256-dimensional embedding instead of an audio waveform, as shown in Fig.~\ref{fig_kd}. The bulk of the TFGridNet architecture computes a 64x65-dimensional representation for each audio chunk, which is then processed by a final convolutional layer followed by inverse-STFT (ISTFT) to compute the output waveform. For the purpose of noisy enrollment, we directly use 64x65-dimensional representation and reduce it using a linear layer to output the 256-dimensional representation for each enrollment audio chunk. We then average the 256-dimensional representations over all enrollment audio chunks to obtain the final 256-dimensional target speaker embedding.
\subsection{Real-time target speech hearing system}\label{sec:realtime}

Now that we have the embeddings for the target speaker, which captures the desired speech traits, our goal is to design a network that can perform target speaker hearing in a way that is both real-time on an embedded CPU and that achieves an end-to-end latency of less than 20~ms. This end-to-end latency is measured as the amount of time for a single sample to pass from the microphone input buffer, through our target speech extraction framework, and then copied into the headphone speaker output buffer, as shown in Fig.~\ref{fig:realtime}(a). We process the input audio in chunks of 8~ms. Moreover, our system utilizes an additional 4~ms of future audio samples to predict the processed output for the current 8~ms chunk. In other words, we must wait for at least 12~ms before we can begin processing the first sample in a particular chunk. Of course, this also means that our algorithm must be designed in such a way that the first sample in the audio chunk does not take into account any information beyond 12~ms into the future, as this information will not be available in practice. 

We design our target speaker hearing network by starting with a state-of-the art speech separation network, namely TFGridNet~\cite{tfgridnet}. However, as this network is non-causal, we adapt the implementation of TFGridNet into a causal version with an algorithmic latency of only 12~ms. To do this, we first remove the group normalization after the first 2D convolution. We then replace the bidirectional sub-band inter-frame LSTM block with a unidirectional LSTM, and fix the unfolding kernel and stride sizes, respectively the hyper-parameters $I$ and $J$, in both recurrent modules to 1. Additionally, instead of computing causal attention using causal masks, inspired by  prior work in \cite{waveformer}, we first unfold the key and value tensors into independent fixed-size chunks using a kernel size of 50 and a stride of 1. We then compute an attention matrix for every chunk between the key tensor and a single-frame query tensor corresponding to the last (rightmost) frame in the chunk, as illustrated in Fig.~\ref{fig:realtime}(b). This attention matrix contains the multiplicative weights of the corresponding frames in the value tensor used to obtain the final output. This ensures that when we predict the output for a single frame, we only attend to the 50 frames that arrive with or before it. Although this limits the duration of time the attention layer looks into the past, this on-device approach is necessary to allow the network to effectively process larger time sequences. We choose an STFT window length of 196 samples, or 12~ms at 16~kHz, and a hop length of  128 samples, or 8~ms at 16~kHz. When computing the ISTFT, we trim the last 4~ms of audio, as these samples will be affected by future chunks during the overlap-and-add operation of the ISTFT. Thus, our resulting output would be 4~ms shorter than the input, which means we obtain an 8~ms output from a 12~ms input.  {For the TSH model, we use the} \verb+asteroid+ ~\cite{asteroid} implementation of STFT.

Once we copy a chunk from the audio buffer into memory, we can begin processing the audio. Since we process audio 8~ms at a time, we need to ensure that each chunk is processed in at most 8~ms on an embedded device, or else incoming chunks begin to queue up, causing the processed output audio to be increasingly delayed. This constraint thus requires us to employ a very stringent processing algorithm to ensure we can keep up with the incoming audio stream. As the original TFGridNet could not meet these runtime requirements on our embedded CPU, we therefore needed to optimize the above model in various ways, we describe below,  to minimize the inference time.

{\bf Caching intermediate outputs.} When processing a stream of continuous chunks, there are numerous values that can be reused to avoid recomputing them. We maintain these values as a list of model state buffers that we pass as an input to the model in addition to the input signal and the target embedding at every inference. For example, we can avoid recomputing the STFT frame of prior chunks by caching these values and re-using them when computing the output of the first 2D convolution layer, as shown in Fig.~\ref{fig:realtime}(c). Likewise, we can store the output of the sequence of GridNet blocks from previous chunks and use them to compute the 2D deconvolution quicker. Additionally, as computing the ISTFT for the current chunk also uses information from  previous chunks, we also need to maintain a buffer for the intermediate outputs of this 2D deconvolution layer. Furthermore, we maintain the hidden and cell states of the temporal unidirectional LSTM  for every GridNet block we have. This allows us to truly make use of the long-term receptive field of the recurrent network. Finally, for every GridNet block, we also maintain the state buffers for previous values of the key and value tensors and concatenate them before unfolding (Fig.~\ref{fig:realtime}(b)). We use these buffers to avoid recomputing the linear projections of previous frames.

{\bf ONNX-specific optimizations.} Our goal is to deploy network using ONNX Runtime~\cite{onnxruntime}. To do this, we re-write certain parts of the network to be more suitable for this setup. First, since we set both $I$ and $J$  to 1, we effectively remove the need for the unfolding layers before the recurrent intra-frame and sub-band modules. Additionally, we can replace all convolution and deconvolution layers, which now have a fixed kernel size and stride of 1, with linear layers that are converted to simpler matrix multiplication kernels when converting to ONNX. We also modify the layer normalization modules to use the native PyTorch implementation, which newer ONNX converters can readily fuse into a single kernel, reducing the overhead of multiple kernel calls.  {Finally, we rewrite the multi-head attention layer, which was implemented as a for-loop to compute the key, query and value tensors for each head, as a single block for each of these tensors, and reshape the output appropriately. This reduces the overall number of nodes in the ONNX graph.}

{\bf Reducing model size.} Instead of using the hyper-parameters suggested in~\cite{tfgridnet}, we choose a hyper-parameter setting that produces a smaller, faster model. In this work, we use $D=64$, $B=3$, $H=64$, $I=1$, $J=1$, $L=4$ and $E=6$. The resulting model has a total of 2.04~million parameters.

To condition the network with the speaker embedding obtained during enrollment, we use a simple linear layer followed by layer normalization to compute a common 64$\times$97 conditioning vector for all time chunks, which we multiply with the latent audio representation between the first and second GridNet blocks.

\subsection{Training  for real-world generalization}\label{sec:training}

We train our target speech hearing system in two steps. We first train the enrollment networks to estimate d-vector embeddings. We then separately train the target speech hearing model while conditioning it with reference d-vector vector embeddings. This approach allows us to use the same target speech hearing model with any enrollment model that can estimate d-vector embeddings. We train these models with a training dataset that considers an accurate representation of real-world use cases of a target speech hearing system. Specifically, we consider variations in speech characteristics, acoustic transformations caused by physical multipath environments, acoustic transformations caused by the human head related transfer function (HRTF)  and diverse background noise. We also consider the effects caused by motion  of the speaker and noise with respect to the listener as an additional finetuning step. Below we explain the dataset details followed by the training process of the enrollment and target speech hearing networks.

\textbf{Synthetic dataset.} Each training sample in our dataset corresponds to an acoustic scene comprised of 2-3 speech samples and background noise. To create an acoustic scene, {we first sample a 5 second background noise sample  and then overlay the target speech and inferring speech at random start positions. For obtaining target and interfering speech,  we randomly select 2-3 speakers from the LibriSpeech dataset \cite{panayotov2015librispeech}, and select a speech sample of length 2-5s for each speaker. The enrollment signals are generated using the same approach as well.} We used the \verb|train-clean-360| component of LibriSpeech dataset that comprises 360 hours of clean speech with 439 and 482 different female and male speakers, respectively. We further select  random  noise samples from WHAMR! dataset \cite{wichern2019wham} comprising a database of audio samples of real-world noisy environments. These audio samples, however, do not contain the effects of real-world indoor environments and human heads, which we found is important for extracting natural-sounding audio.

\textbf{Accounting for multipath and HRTF.} To account for these effects, we convolve each of the speech samples and background noise with a binaural room-impulse-response (BRIR) that captures the acoustic transformations caused by a room as well as a user's head and torso. Let $h_{r, \theta,\phi}$ be a BRIR corresponding to the room and  subject combination $r$, at azimuthal angle $\theta$ and polar angle $\phi$ with respect to the subject's head. Let $S_0(t) \in R$ and $S_1(t) \in R$ be two mono clean speech mixtures sampled from the LibriSpeech dataset, and $V(t)$ be noise sampled from the WHAMR! dataset. Then the binaural acoustic scene $x(t) \in R^2$ for this  source mixture could be computed as:
$$ x(t) = S_0(t) * h_{r, \theta_0,\phi_0} + S_1(t) * h_{r, \theta_1,\phi_1} + V(t) * h_{r, \theta_v,\phi_v}$$

It is to be noted that in each acoustic scene, room and subject configuration $r$, remains the same for all sources, but the angles with respect to the listener are arbitrary. To improve robustness to variations in rooms and subject, we aggregate BRIRs from 4 different datasets: CIPIC \cite{CIPIC}, RRBRIR \cite{rrbrir}, ASH-Listening-Set \cite{ShanonPearce} and CATTRIR \cite{CATT_RIR}. Of these, CIPIC dataset only comprises of impulse responses measured in an anechoic chamber and as a result, is devoid of any room characteristics. Combined, these datasets provided us with a total of 92 different room and subject configurations.


\textbf{Training.} To train the  enrollment networks, we first generate the component speech utterances, {as described above, with the constraint that target speaker's azimuthal angle $\theta_{0} \sim \frac{\pi}{2}$. {We train the beamformer-based enrollment network to predict target speech (Fig. \ref{beamformer}) with SNR loss. We train knowledge-distillation-based enrollment network to predict the d-vector embedding of the target speech (Fig. \ref{fig_kd}) with cosine-similarity loss.}

\begin{figure*}[t!]
\centering
\includegraphics[width=0.95\textwidth]{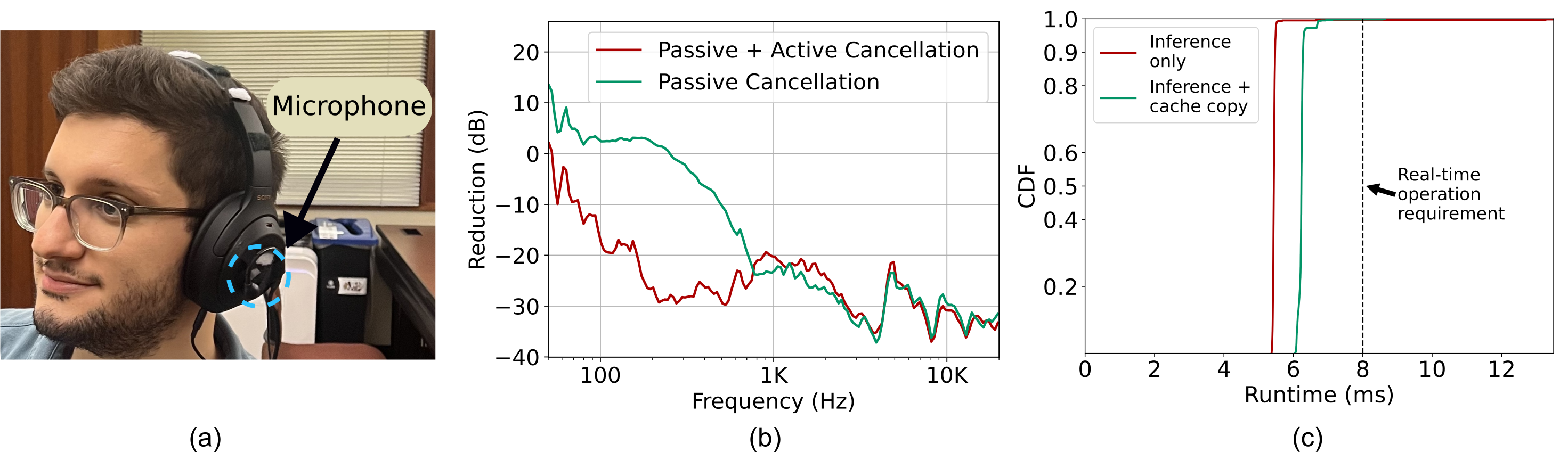}
\vskip -0.13in
\caption{{\bf End-to-end target speech hearing system with noise cancellation.} (a) A user wearing our proposed hardware system. (b) Noise isolation performance of a pair of Sony WH-1000XM4 headphones with and without active noise cancellation across frequencies. The 
larger values at low frequencies  are due to the in-ear
microphones picking up the wearer’s blood pulse. (c) CDF plot of the model inference time, with and without the cache buffer copy from the outputs to the inputs. }
\label{fig:system_anc_rt}
\end{figure*}

To train the target speech hearing (TSH)  network,  we also sample a random speech corresponding to the target speaker and convolve it with a BRIR corresponding to the same room and subject configuration. {We input the TSH model with the acoustic scene and the d-vector embedding computing on the sample speech. We then optimize the TSH network, $\mathcal{T}$, to minimize the signal-to-noise ratio \cite{https://doi.org/10.48550/arxiv.1811.02508} (SNR) loss between the estimated target speech and the ground-truth: -SNR($\hat{s}_0(t), s_0(t)$).}



\textbf{Finetuning for  motion, error in the enrollment angle and real-world noise characteristics.} In the dataset setup described above, we assumed a constant azimuthal angle for each source as time progressed. This means that sources are stationary with respect to the listener's orientation, and the enrollment angle is close to $\frac{\pi}{2}$ and does not change with time. These assumptions, however, are not true in the real world as sources could be moving, or there might be a rotation in the listener's head resulting in significant relative angular velocities.

We handle relative motion and time-varying error in the enrollment angle with an additional finetuning step. During finetuning, we make the azimuthal and polar angle time-varying. We simulate motion by generating an array of positions over time with a finite time step of 25~ms. For enrollment, we assume that at each time step, both the enrollment azimuth and enrollment polar angles are uniformly random and lie in the range $[\frac{\pi}{2} - \frac{\pi}{10}, \frac{\pi}{2} + \frac{\pi}{10}]$, accounting for a maximum error of 18 degrees. For the rest of the sources -- interfering sources in the enrollment acoustic scene, and interfering as well as target sources in the input to the TSH model -- we generate a random speaker motion by randomly triggering speaker motion events at each time step with a probability of 0.025. When a speaker motion event is triggered, we sample a pair of angular velocities along the polar and azimuthal directions with magnitudes uniformly distributed in the range $[\frac{\pi}{6}, \frac{\pi}{2}]$~rad/s. The speaker moves along this direction for a random duration uniformly sampled from $[0.1, 1]$~s. During this time, we do not consider any other motion events. This creates trajectories where the speaker may be stationary for some time intervals and sporadically move with different velocities within the same audio clip.  Assuming such time-varying trajectories, the computation of an enrollment scene could be written as:
$e(t') = S_{0}(t') * h_{r, \theta_{0}(t'),\phi_{0}(t')} + S_{e1}(t') * h_{r, \theta_{e1}(t'),\phi_{e1}(t')} + V_e(t') * h_{r, \theta_{ve}(t'),\phi_{ve}(t')} $, 
where, $\theta_{0}(t'), \phi_{0}(t') \in [\frac{\pi}{2} - \frac{\pi}{10}, \frac{\pi}{2} + \frac{\pi}{10}]$. And the input scene can be computed as:
$$ x(t) = S_0(t) * h_{r, \theta_0(t),\phi_0(t)} + S_1(t) * h_{r, \theta_1(t),\phi_1(t)} + V(t) * h_{r, \theta_v(t),\phi_v(t)}$$

Since the BRIR datasets typically provide an impulse response at discrete points in space, it is not possible to directly perform the computation described in the expressions above. To achieve such a trajectory simulation with available BRIR datasets, we employ a nearest neighbor approximation of BRIRs -- we select BRIR in the dataset that is closest to the desired azimuth and polar angle in time. We use Steam Audio SDK \cite{steamaudio-sdk} to perform this motion trajectory simulation. During the finetuning step, we use all four BRIR datasets we described above, but perform motion simulation only with CIPIC database \cite{CIPIC}. We do this because CIPIC provides BRIRs that are reasonably dense across all azimuth and polar angles, while the rest of BRIR datasets sparsely vary azimuthal angle only with fixed elevation.

{Finally, to allow the model to learn common noise characteristics found in the real world such as microphone thermal noise and constant noise from heating, ventilation and air conditioning systems, we also train the model with randomly scaled white, pink and brown noise components. Specifically, during training, we augment the mixture signal with a white noise signal having a standard deviation uniformly chosen from the range $[0, 0.002)$.} As for the pink and brown noise, we use the \verb+powerlaw_psd_gaussian+ function from the Python \verb+colorednoise+ library to generate these noise signals. We scale the pink and brown noise signals with  different scale factors each sampled uniformly from $[0, 0.05)$.


\section{Implementation and Evaluation}

We first describe our end-to-end integration with noise-canceling headsets. We then describe our   in-the-wild evaluation with previously unseen speakers and acoustic environments. Next, we describe our user study comparing different enrollment interfaces. Finally, we present benchmarks for the various models. 

\subsection{End-to-end system with noise-cancellation}\label{sec:noisecanceling}

\begin{figure}[t!]
\centering
\includegraphics[width=\columnwidth]{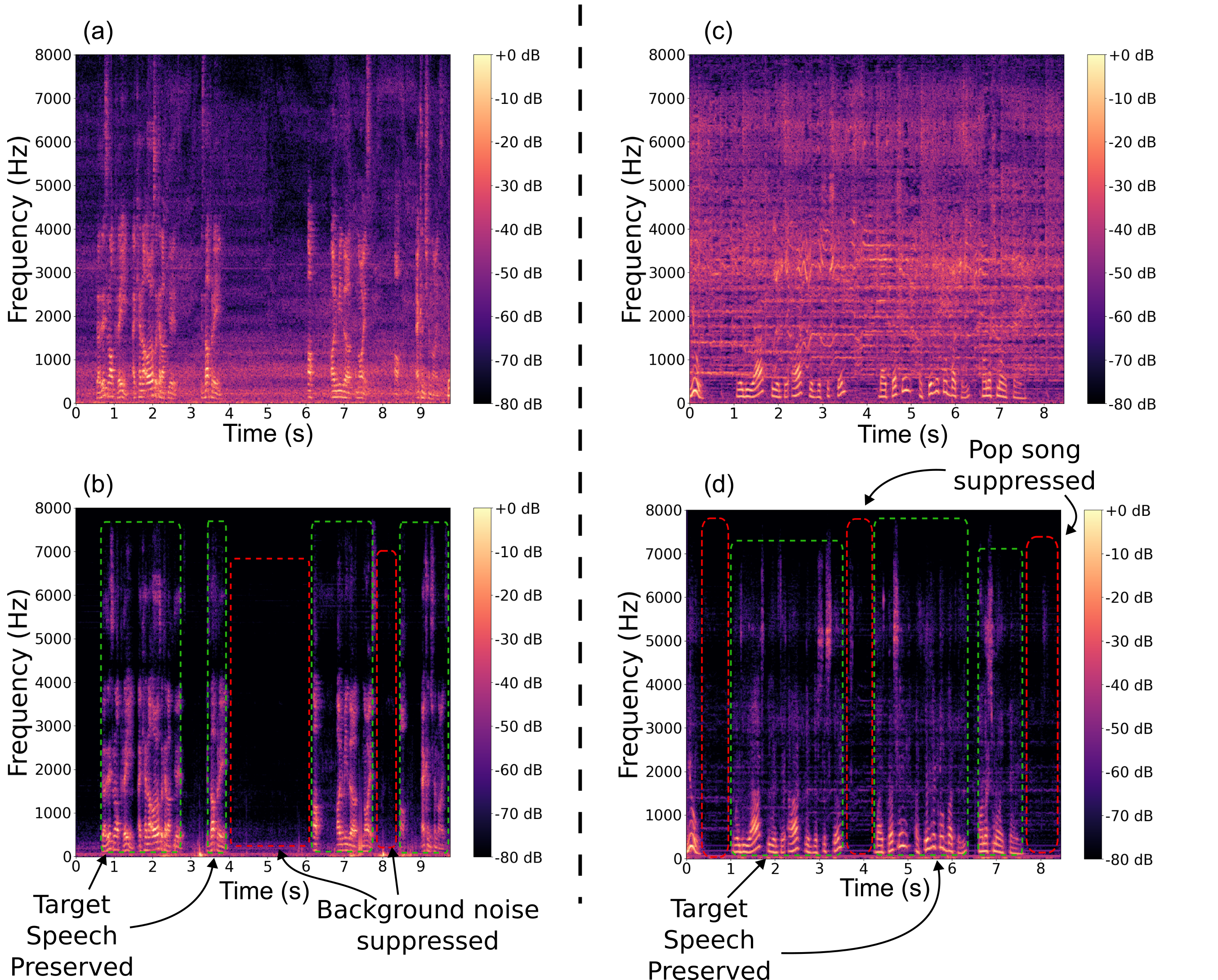}
\caption{{\bf Recordings from microphones placed both outside and inside the earcup, with a participant wearing the hardware.} In (a), the spectrogram of the speech from a target speaker is obscured by the sounds of an airplane engine, which our system can effectively remove, as shown in the spectrogram of the recorded audio inside the earcup shown in (b). We see a similar pattern in (c), where the background song and target speech heard outside the headphone cups are processed and played back to the user, for whom the background song gets suppressed and the speech preserved, as shown by the audio recorded inside the earcups shown in (d). }
\vskip -0.2in
\label{fig:spectra}
\end{figure}


\textbf{Hardware prototype.} We design our hardware using the Sony WH-1000XM4 noise-cancelling headset. Since we need binaural acoustic data, we attach a pair of binaural microphones (Sonic Presence SP15C) to the exterior side of the earcups of the noise-canceling headset as shown in Fig.~\ref{fig:system_anc_rt}(a). The incoming binaural audio is processed on an embedded CPU and is played back to the user using the noise-canceling headset.  Fig.~\ref{fig:system_anc_rt}(b) shows the attenuation achieved by the noise-canceling headsets across different frequencies, with and without active noise cancellation. In these experiments, we obtain recordings from a pair of microphones fitted into a human user's ears as a nearby loudspeaker plays a 20~Hz-22~kHz linear frequency sweep.  While wearing the headphones, we notice that the earcups cause the microphone to pick up an audio signal that synchronizes with the wearer's blood pulse, which produce the  spurious peaks  at the lower frequencies (<100~Hz).


\textbf{Runtime evaluation.} We connect the binaural microphones  to an Orange Pi 5B using a USB cable. This allows the Orange Pi to read audio chunks from the microphones, which we process on-device and play back to the headset connected via the audio jack. We deploy our neural network on the embedded device by converting the PyTorch model into an ONNX model using a nightly PyTorch version (2.1.0.dev20230713+cu118) and we use the python package \verb+onnxsim+ to simplify the resulting ONNX model. We run inference using the ONNX Runtime version 1.16.0. To evaluate our model runtime, we use the ONNX Runtime \verb+perftest+ tool to run 1000 successive inference operations with the model, and we observe that the model inference time, which is 5.47~ms on average, is lower than our chunk size of 8~ms.  This means that our model can process an audio chunk before the next chunk arrives, and hence, it satisfies our real-time requirement. 

In practice, since our model makes significant use of cached values, which we must provide as an input  every time we run inference, we must also take into account the time it takes to copy these updated cache buffers from the model output to the model input. When we include these buffer copy operations in our runtime measurements, we see that although the overall runtime increases, it is still within the constraints for real-time operation with a runtime of around 6.24~ms, as shown in Fig.~\ref{fig:system_anc_rt}(c). This gives us a total end-to-end latency of 18.24~ms.


\begin{figure}[t!]
    \centering
    \includegraphics[width=0.99\columnwidth]{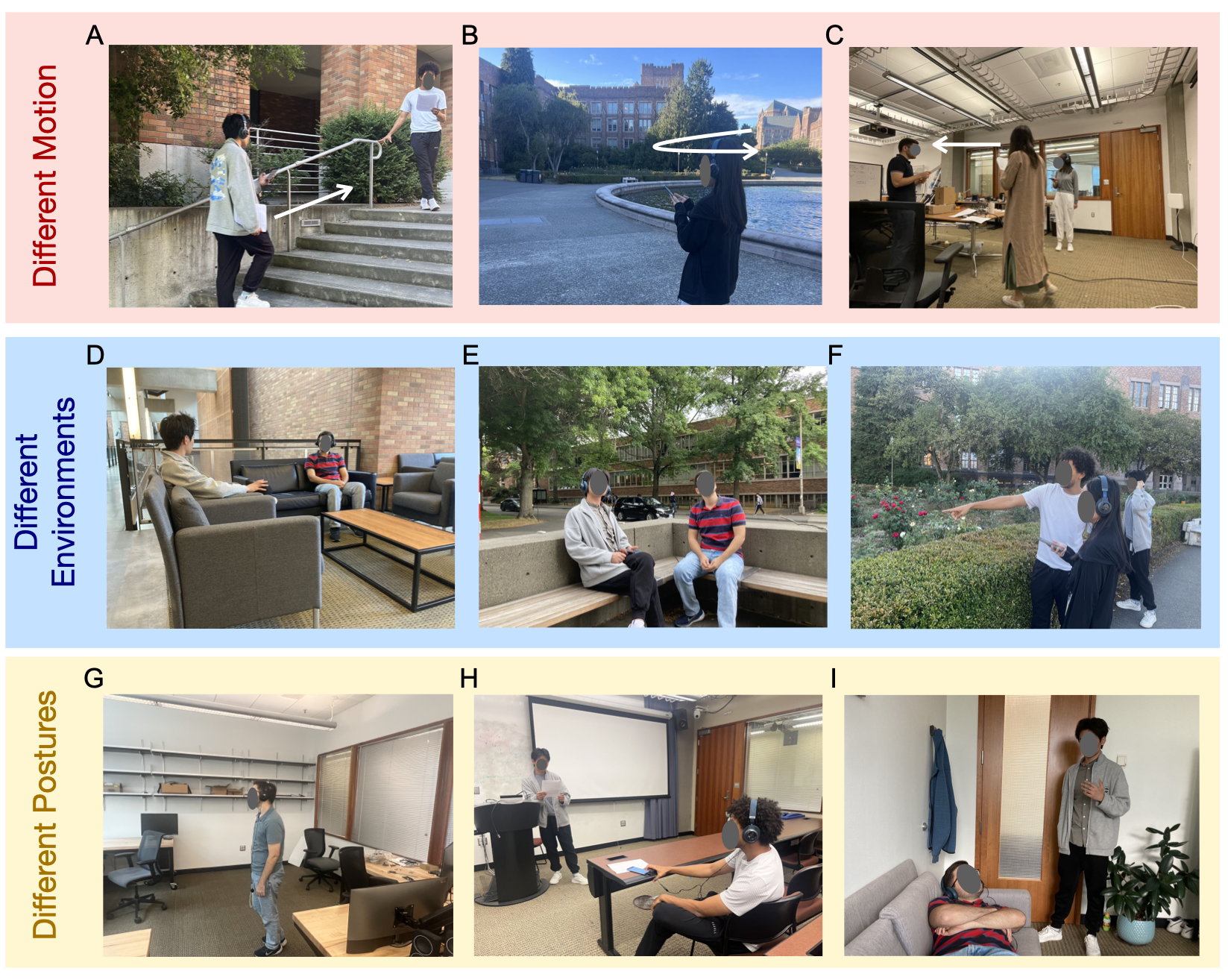}
    \vskip -0.15in
    \caption{{\bf In-the-wild scenarios.} Different scenarios in the real-world evaluation of our system.}
    \vskip -0.2in
    \label{fig:itw}
\end{figure}

\textbf{End-to-end demonstration.} We demonstrate our end-to-end system and ask participants to wear and use it. We enable active noise canceling on the headphones and run our target speaker hearing pipeline after the user enrolls their target speaker. For these experiments, we also place an additional pair of microphones inside the headphone's earcups to record the exact sound that the user will listen to during operation. This includes any residual noise from the imperfect noise cancellation, as well as the amplification of target speakers from our target speaker hearing playback signal. We show the resulting spectrogram representation for two examples in Fig.~\ref{fig:spectra}.  {For brevity, we only show one audio channel.}

In the first example, the user is trying to listen to a target human speaker in the presence of airplane engine noise playing over a nearby smartphone speaker. This noise is clearly visible in Fig.~\ref{fig:spectra}(a), which was recorded from the outer microphone. However, when we probe the sound at the inner microphone (Fig.~\ref{fig:spectra}(b)), we noticed a significant attenuation of this background noise due to the headphones'  active noise-canceling abilities. And yet, because of our target speaker hearing system, the target speaker is still very clearly audible at the inner microphone. We show in the second example that this behavior can even be seen for other, more dynamic and loud background sounds. Specifically, we show in Fig.~\ref{fig:spectra}(c) that even when the user tries to listen to the target speaker in the presence of music, in this case a loud pop song, our end-to-end system have  an impressive ability to suppress the unwanted music and the singer while preserving the sound of the target speaker  (Fig.~\ref{fig:spectra}(d)).

\subsection{In-the-wild generalization}

We  evaluate our hardware  in previously unseen indoor and outdoor environments, with participants  who are not in the training data. 
 We recruited  8 individuals (5 male, 3 female) to  collect data in different in-the-wild scenarios using our   hardware. We ask 3 participants at a time to collect noisy enrollment signals, as well as noisy real-world mixture audio, in different acoustic environments while  they  read  random text. Among the three participants, one of them is designated as the wearer, while the other two are the speakers.   To collect a noisy enrollment for a given target, the wearer looks at the target speaker as they read a text. {This mimics the "Look Once" phase in the real-world use, where the listener would look at the target speaker.} As the target speaker reads the text, background sounds and, in all but one case which had significant noise, speech from the other speaker make this enrollment signal noisy. 


We record multiple noisy binaural audio clips while the target speaker reads a different text in the presence of other environmental  sounds and speakers. Unlike the noisy enrollment signal, there is little control over these recordings, as the wearer and target speaker are free to move around and/or rotate their head (Fig.~\ref{fig:itw}A-C). These recordings were also collected in  different acoustic environments, including living spaces, busy streets, and in nature (Fig.~\ref{fig:itw}D-F). They also contained settings where the listeners were in different postures, such as standing, sitting and laying down (Fig.~\ref{fig:itw}G-I).



{\bf Evaluation procedure.} Since the target speaker is speaking in the presence of interfering speakers and unknown noise, it is difficult to obtain the ground truth audio signal for our target speakers in the real world. So, we cannot rely on objective metrics to evaluate the system performance. Instead, we design a listening survey to allow human participants to rate the performance of our two enrollment methods on 15 different target speaker scenarios from the in-the-wild dataset we collected. To do this, we recruited 21 participants (13 male and 8 female with an average age of 30.4 years) to take our survey and give their opinion on our target speaker hearing system to obtain a mean opinion score (MOS). To do this, for each scenario, we first ask the users to try listening to a  5-second clean  signal of the target speaker reading text collected in a quiet room. We then ask the participants to listen to the target speaker in 3 distinct audio clips:
 1) the original, noisy recording of the target speech with interferers, 2) the output of our target speaker hearing network using the noisy knowledge distillation embeddings, and 3) the output of our target speaker hearing network using the noisy beamformer embeddings.   The  3  clips are presented in random order.

After listening to the  mixture and model output clips in a random order, we ask the participants to rate the target speech extraction quality by asking them the following questions:

\begin{enumerate}
    \item \textbf{Noise suppression}: \textit{How INTRUSIVE/NOTICEABLE were the INTERFERING SPEAKERS and BACKGROUND NOISES? 1 - Very intrusive, 2 - Somewhat intrusive, 3 - Noticeable, but not intrusive, 4 - Slightly noticeable, 5 - Not noticeable}
    \item \textbf{Overall MOS}: \textit{If the goal is to focus on this target speaker, how was your OVERALL experience? 1 - Bad, 2 - Poor, 3 - Fair, 4 - Good, 5 - Excellent}
\end{enumerate}

\begin{figure}[t]
    \centering
    \includegraphics[width=0.95\columnwidth]{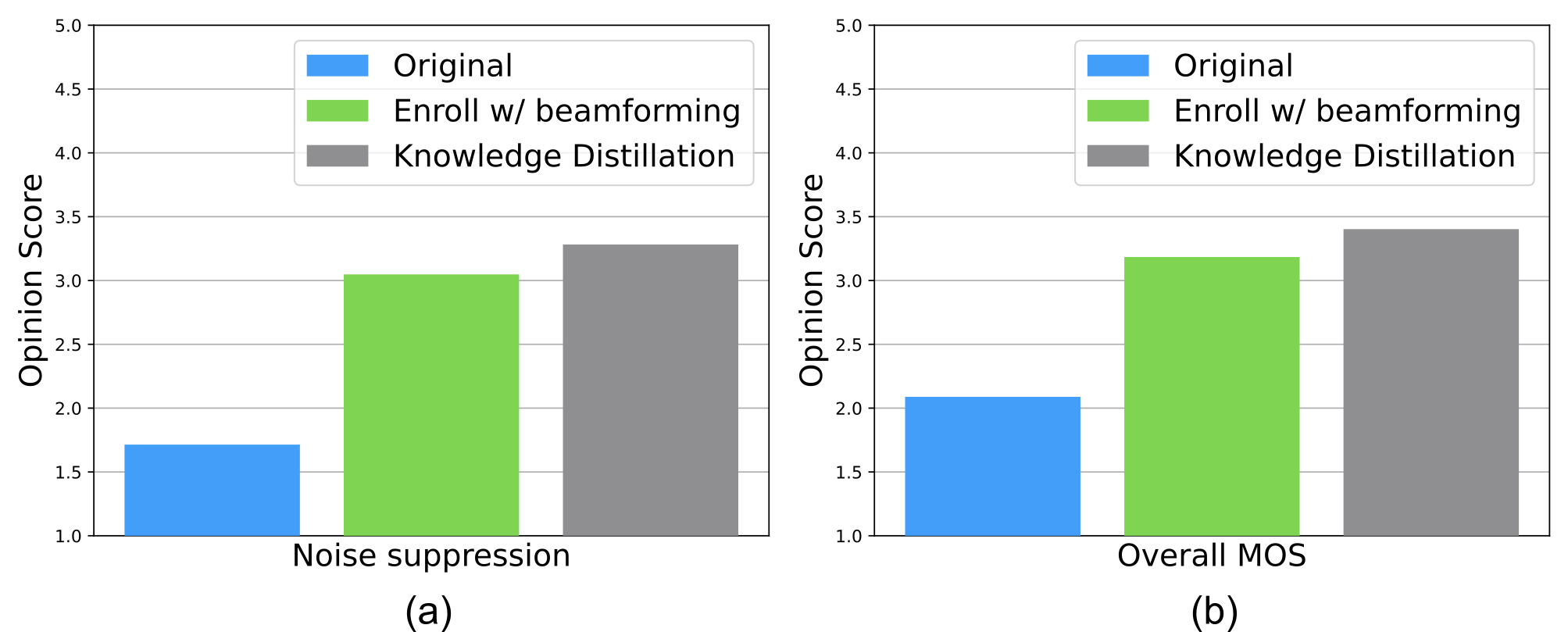}
    \vskip -0.15in
    \caption{{\bf Subjective in-the-wild evaluations.} (a) Mean opinion score for the noise suppression quality reported for the raw audio signal and the output using our two enrollment networks, and (b) overall reported mean opinion score. {Paired t-tests between knowledge distillation and beamforming approaches resulted in $p$-values < 0.001.} }
    \vskip -0.1in
\label{fig:itw_results}
\end{figure}

\begin{figure}[t]
    \centering
    \includegraphics[width=0.99\columnwidth]{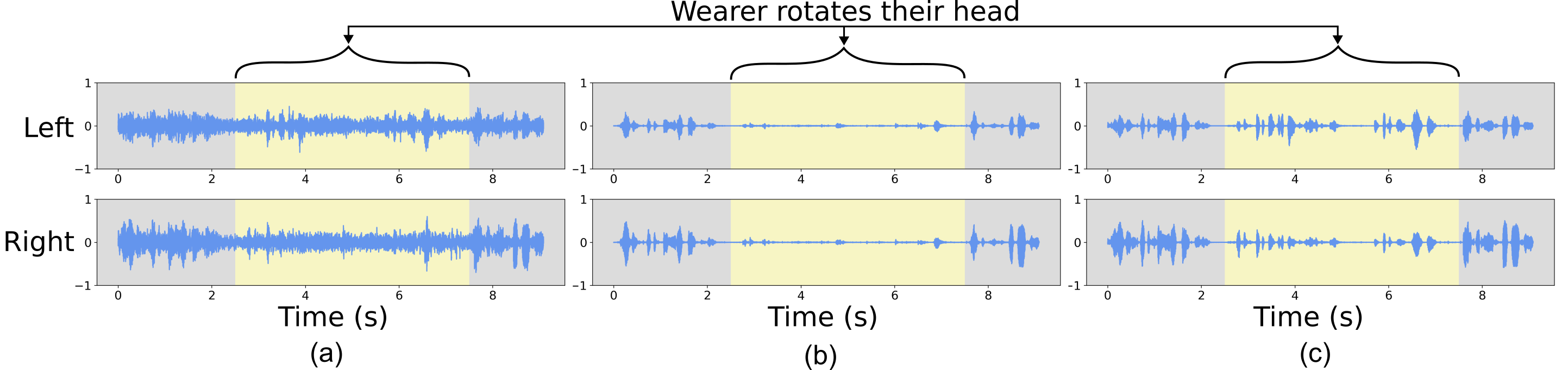}
    \vskip -0.15in
    \caption{{\bf Waveforms of a  real-world noisy recording of a target speaker with a mobile wearer.} In (a), we see the original recording without any processing, (b) shows the output of the target speaker hearing network without fine-tuning with moving sources, while (c) shows the output after fine-tuning.}
    \vskip -0.1in
    \label{fig:itw_motion}
\end{figure}

{\bf Results.} Fig.~\ref{fig:itw_results}  shows that our system can greatly suppress the background sounds and interfering speakers, as evidenced by the fact that that our beamforming  and knowledge distillation enrollment networks increased the mean opinion score for the noise suppression task from 1.71 to 3.05  for the beamformer enrollment method and 3.28  for the  knowledge distillation method (Fig.~\ref{fig:itw_results}(a)).  Our target speaker hearing framework was also able to improve the overall mean opinion score from 2.09  to 3.18  and 3.4, respectively 
 (Fig.~\ref{fig:itw_results}(b)).

\begin{figure}[t!]
\vskip -0.1in
    \centering
    \begin{subfigure}[t]{0.49\columnwidth}
        \centering
        \includegraphics[width=1.13\columnwidth]{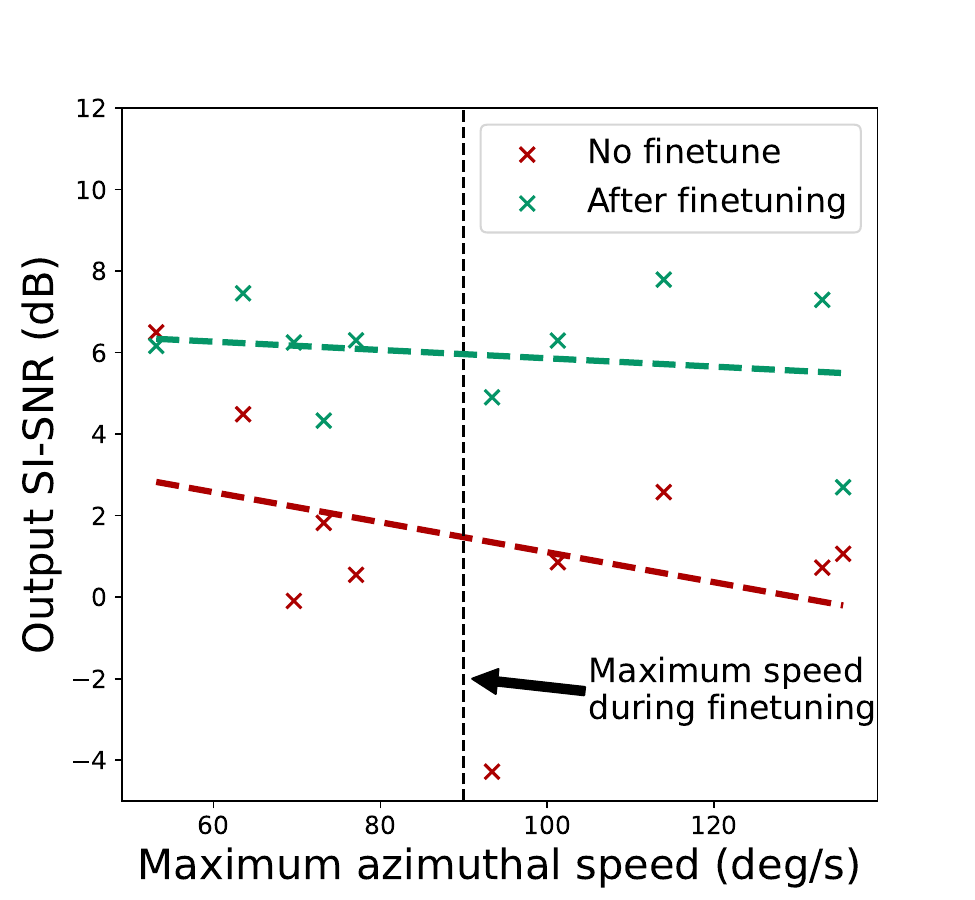}
        \caption{}
        \label{fig:rw_azi}
    \end{subfigure}
    \begin{subfigure}[t]{0.49\columnwidth}
        \centering
        \includegraphics[width=1.13\columnwidth]{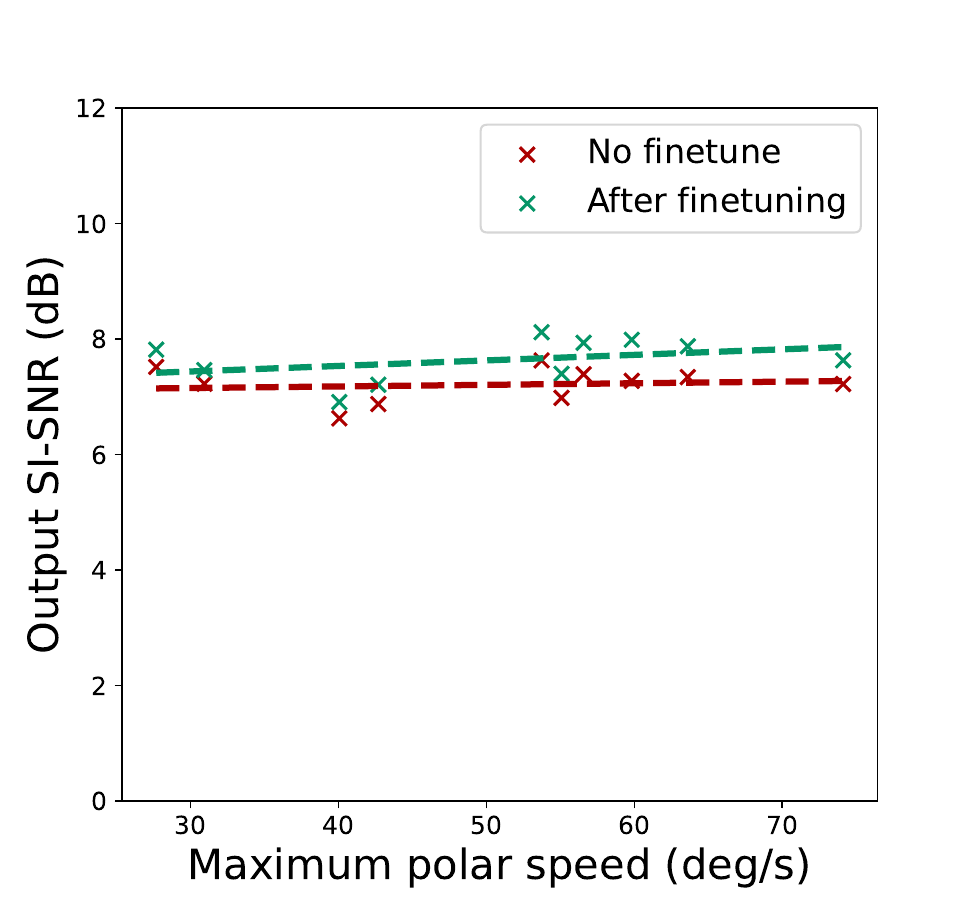}
        \caption{}
        \label{fig:rw_polar}
    \end{subfigure}
    \vskip -0.15in
    \caption{{ Real-world performance comparison with and without fine-tuning. The maximum angular rotation is reported as the maximum absolute  average speed over 100~ms.}}
    \vskip -0.15in
\label{analysis}
\end{figure}

The results show a consistency between our  objective benchmark results in~\xref{sec:benchmark} and real human evaluators, in that the embedding network trained with knowledge distillation outperforms the beamforming network in both cases. 
These evaluations also show that our system can generalize well to real-world environments with real human wearers. One particularly important scenario is to be able to adapt to sudden and rapid changes in the target speaker's position due to the wearer rotating their head. Indeed, many of the participants in our study often rotated their heads, looking at different speakers and objects in their surroundings. As a result, vital spatial cues  could vary widely with time. Such scenarios highlight the importance of fine-tuning our neural networks with audio samples that contain moving speakers. For example, in Fig.~\ref{fig:itw_motion}, we show a real-world example where the listener was noticeably rotating their head while recording throughout the highlighted region. This can be clearly seen by observing the level differences between the left and right channels, as the target speaker becomes more prominent in the left channel in that time frame. As a result, the target speaker appears to move relative to the listener. Without fine-tuning, our network incorrectly suppresses the target speaker during this time. However, when we fine-tune the network with our mobility technique in \xref{sec:training}, we see that it can correctly pick the target speaker even when the user turns their head. 

{To further examine this behavior, we conduct an experiment with motion in the real world. Our goal here is to evaluate the model's ability to adapt to the wearer's head motion. We play 6 seconds of speech from a loudspeaker and record audio from a participant wearing our  system in an office room with HVAC and other ambient noise. As the speech is played, the participant rotates their head several times per trial with different speeds, pausing briefly between successive rotations. We obtain  recordings where the wearer rotates exclusively in the azimuthal direction, and  recordings where they  rotate exclusively the polar direction. We use a gyroscope to record the angular velocity during each trial. 
To evaluate the model, we use a different recording of this same speaker as an embedding signal and process the recorded audio. Since there is no other interfering speaker in this experiment (computing SI-SNR in real-world mixtures is challenging with  interference), a well-trained model is expected to preserve the original speech regardless of the user rotation. As we see in Fig.~\ref{fig:rw_azi}, when the participant rotates their head in the azimuthal direction (horizontally), there is a clear drop in performance without fine-tuning. Additionally, we see that this drop becomes worse as the rotation speed increases. After finetuning, the model is able to generalize to such  motion, however the variance in the performance increases for rotation speeds outside the training bounds. On the other hand, we see that the model generalizes to rotations in the polar direction (vertically) even without finetuning, albeit with a slight performance decline (Fig.~\ref{fig:rw_polar}). This is likely because when the wearer moves their head vertically, the position of the microphones do not change considerably. This analysis suggests that the model is leveraging inter-channel differences. Therefore, it is important to use moving sources during training for the system to better adapt to real-world use cases.
 }

\begin{figure}[t]
    \centering
    \includegraphics[width=0.97\columnwidth]{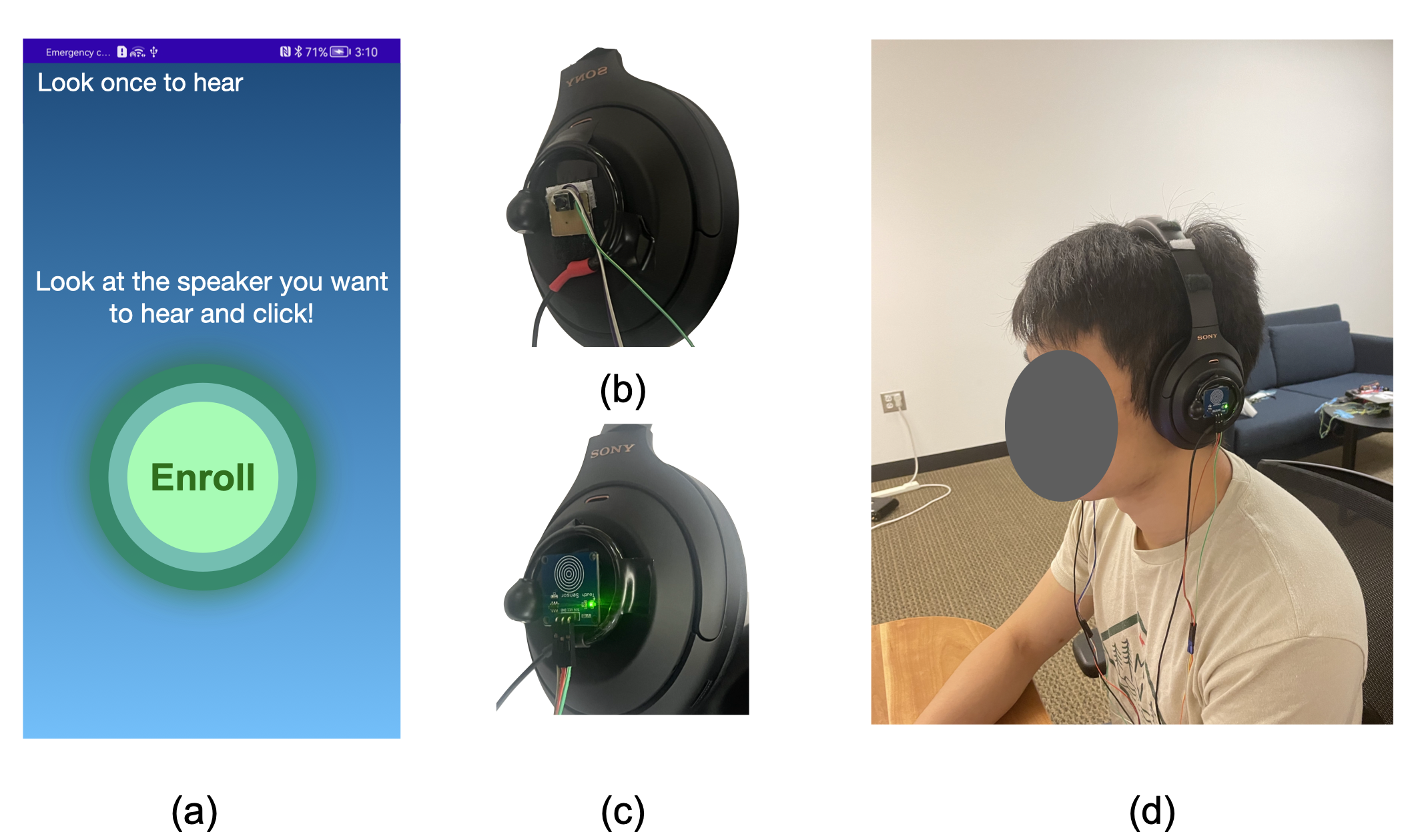}
    \vskip -0.15in
    \caption{{\bf Proposed  interfaces.}  (a) A smartphone app, (b) a push button, and (c) a touch pad. In (d), we see a participant wearing our prototype while conducting the user study.}
    \vskip -0.1in
    \label{fig:ui}
\end{figure}

\subsection{Enrollment interface user study}


We investigate two main questions: 1) What interface should the user interact with when they want to enroll the target speaker?, and 2) what enrollment duration do  users find acceptable for such a system?  As shown in Fig.~\ref{fig:ui}, we integrate into our prototype  with three different interfaces through which users can communicate their intention of enrolling a target speaker: 1) a virtual button on a smartphone application, 2) a push button on the headphone, and 3) a touch pad on the headphone. We evaluate four  different possible enrollment durations: 2.5~s, 5~s, 7.5~s and 10~s.

\textbf{Comparing user interfaces.}
We conducted a user study with 9 participants, where each participant first wore our device and sat on a chair as shown in Fig.~\ref{fig:ui}(d). We placed a loudspeaker to the side of the wearer, which played a mixture of human speech from the LibriSpeech dataset and generic vacuum noise. A person sitting in front of the participant would then read a random text. We then asked the participants to use each of the three interfaces to signal to the system their intention to enroll the target speaker in front of them, while suppressing the interfering sounds emitted by the loudspeaker. When the users correctly interacted with the device to start enrollment, the headset would play a voice saying "Enrollment start". While enrolling the target speaker, the participants are asked to keep their head facing the target speaker, until the enrollment duration has passed and the enrollment is completely recorded, at which point a  voice is played over the headphones, saying "Finished".  The enrollment duration for all three interfaces was set to 5 seconds. After interacting with the three interfaces, we asked each participant to rate the three interfaces from 1-5 based on how likely they are to use that interface for this interaction. The results are shown in  Fig.~\ref{fig:score}(a). Most participants showed a strong preference for the push button because of its good haptic feedback. All the participants showed the least preference for the smartphone   since  it required the extra complexity of looking at the smartphone screen while simultaneously trying to face the target speaker.

\begin{figure}[t!]
    \centering
    \includegraphics[width=0.99\columnwidth]{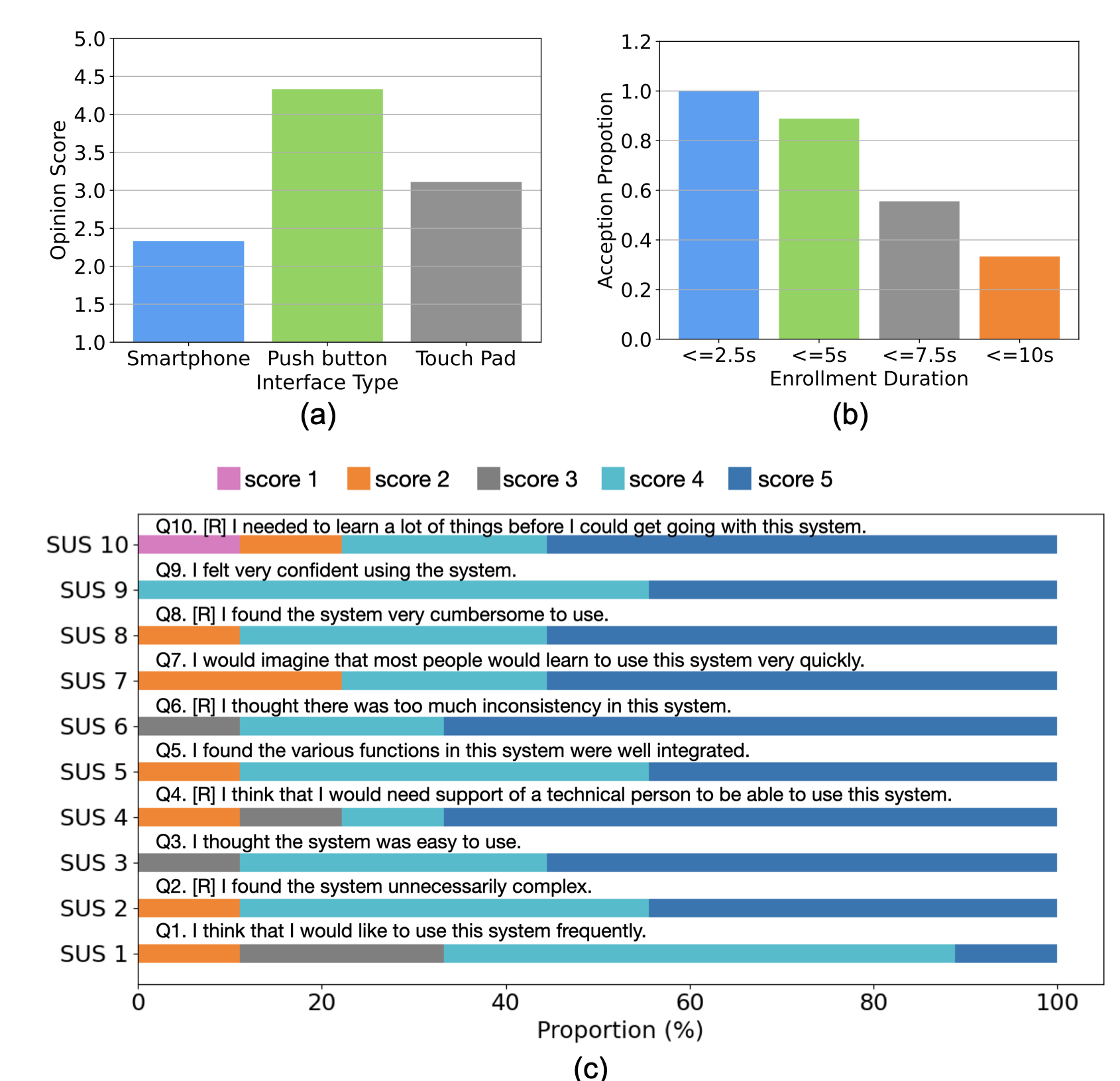}
    \vskip -0.15in
    \caption{{\bf Results of our user study. } (a) shows that the push button was the most preferred interaction method, while the smartphone app was the least preferred option.  (b) shows the participant preferences with different enrollment duration.  (c)  shows the  results of the SUS questionnaire, where we reverse the scale of negatively worded statements (Q2, Q4, Q6, Q8 and Q10) for easier visualization.}
    \vskip -0.2in
    \label{fig:score}
\end{figure}

\begin{table*}[t!]
    \centering
    \caption{Benchmarking results on the generated test set. Proposed noisy enrollment methods are evaluated with 3 different audio/speech processing architectures. Performance with clean enrollments is also provided for reference.}
    \begin{tabular}{lccccc}
    \toprule
    Enrollment & d-vector &  Real-time & SI-SNRi (dB) & Params (M) & MACs (GMAC)\\
    network & similarity & TSH backbone& & & \\
    \midrule
     Clean &  1.0 & Streaming TFGridNet & 7.40 & 2.04 & 4.63 \\
           &      & Waveformer & 4.94 & 1.6 & 2.43 \\
           &      & DCCRN & 6.71 & 5.54 & 6.6 \\
    \midrule
     Beamformer &  0.74 & Streaming TFGridNet & 4.53 & & \\
           &      & Waveformer & 2.34 & " & " \\
           &      & DCCRN & 4.34 & & \\
    \midrule
     Knowledge distillation &  0.85 & Streaming TFGridNet & 7.01 & & \\
           &      & Waveformer & 4.63 & " & " \\
           &      & DCCRN & 6.16 & & \\
    \bottomrule
    \end{tabular}
    \vskip -0.1in
    \label{benchmark}
\end{table*}

\textbf{Evaluating the enrollment duration.} We then asked the participants to use their favorite interface from the previous study to explore their perspectives on a reasonable enrollment duration. Specifically, each participant performed enrollments with 4 different durations:  2.5s, 5s, 7.5s and 10s. The results in  Fig.~\ref{fig:score}(b) show that 89\% of the participants thought that 5 seconds was an acceptable duration for the enrollment period.

\textbf{Qualitative Results.} Next, we asked each of the participants to fill out a System Usability Scale (SUS) questionnaire, which was developed in~\cite{SUS}. The overall score was $80.8\pm16.7$, which suggests generally positive feedback on usability. The SUS results for each question are shown in  Fig.~\ref{fig:score}(c), which correlates with being highly usable and acceptable by users, according to Bangor's empirical evaluation~\cite{Bangor}. Finally, in addition to the previous studies, we also ask the subjective question: "Where do you see yourself using such a system?". Five of the participants mentioned using them in crowded scenarios and expressed similar applications to the response of one of the participants: "I'd like to use it in large social gatherings like conferences and lectures. I want to just  talk with a specific people without being distracted by others or loud background noise". Five of the participants also mentioned that they were willing to use it in common public locations such as cafes, restaurants, on the street, karaoke and in large parties. Furthermore, one participant proposed that this technology might be useful for hearing aids. While all participants gave positive feedback and proposed useful potential applications, two  of them also raised some limitations. One participant said, "In the real-world, I would also  want to focus on a group of people instead of only one person". Another participant said "I think the headphone form factor is a bit obtrusive. A wireless earbud form-factor would be more socially acceptable."

\subsection{Benchmarking the models} \label{sec:benchmark}


We quantitatively compare our two  noisy enrollment methods described in~\xref{sec:enrollment}.  We also compare the quality of target speech with that extracted with embeddings computed on clean enrollments.


For the real-time target speech-hearing aspect of the system, we consider three  models: our embedded real-time implementation of TFGridNet \cite{tfgridnet}, Waveformer \cite{waveformer} and DCCRN \cite{hu2020dccrn}. These have been developed for speech separation, target sound extraction and speech enhancement, respectively.  Waveformer and DCCRN are causal real-time models suitable for this task.  
 {We use network variants with the default algorithmic latency proposed in the respective architectures of 13~ms and 37.5~ms for Waveformer and DCCRN respectively, while for our causal TFGridNet implementation, we use an algorithmic latency of 12 ms.}


This results in a total of nine combinations -- 3 enrollment methods with 3 TSH backbones. We evaluate all nine combinations with 10000 binaural mixture/noisy enrollment pairs generated as described in \xref{sec:training}.  Table \ref{benchmark} reports the average target speaker's signal quality improvement with respect to the mixture in terms of scale-invariant signal-to-noise ratio improvement (SI-SNRi). Assuming speaker embeddings computed from clean enrollment signals as a reference, we also report the cosine similarity between reference embeddings and embedding computed by both noisy enrollment methods in the second column.  The last two columns show the size of the architecture in terms of parameter count as well as multiply-and-accumulate per second (MACs/s).  Among these architectures, we find that TFGridNet offers the best performance while also consuming sufficiently low operations and latency.



\begin{table}[b!]
\vskip -0.15in
    \centering
    \caption{{Runtime comparison  of various optimizations on a TFGridNet with a similar parameter configuration. We report the average runtime over 1000 forward passes. To obtain runtime measurements, for the PyTorch model, we use the PyTorch v2.1.1 and the PyTorch Profiler. We use the knowledge distillation embedding network for SI-SNRi evaluation.}}
    {\footnotesize
    \begin{tabular}{lccc}
    \toprule
    Model & SI-SNRi (dB) & Runtime  (ms) \\
    \midrule
     Streaming TFGridNet w/ caching on  PyTorch & 7.12 & 607.93\\
     Streaming TFGridNet w/ caching on  ONNX & " & 9.28\\
    Optimized Streaming TFGridNet with ONNX & 7.01 & 5.47\\ 
    \bottomrule
    \end{tabular}}
    \label{runtime_comparison}
\end{table}

\begin{figure*}[t]
    \centering
    \begin{subfigure}[t]{0.45\textwidth}
        \centering
        \includegraphics[height=1.1in]{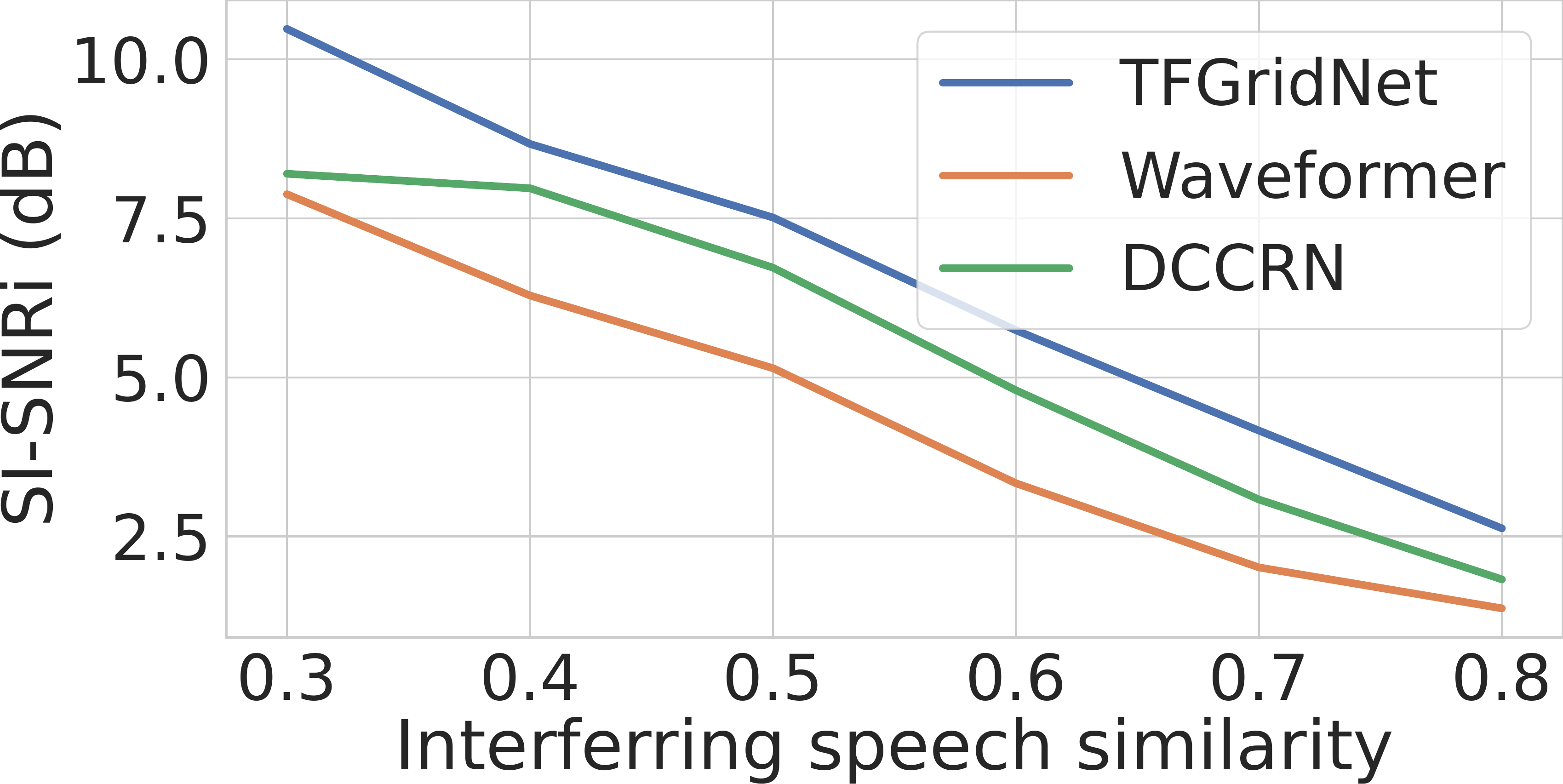}
        \caption{}
        \label{intefer_speech}
    \end{subfigure}%
    ~
    \begin{subfigure}[t]{0.45\textwidth}
        \centering
        \includegraphics[height=1.1in]{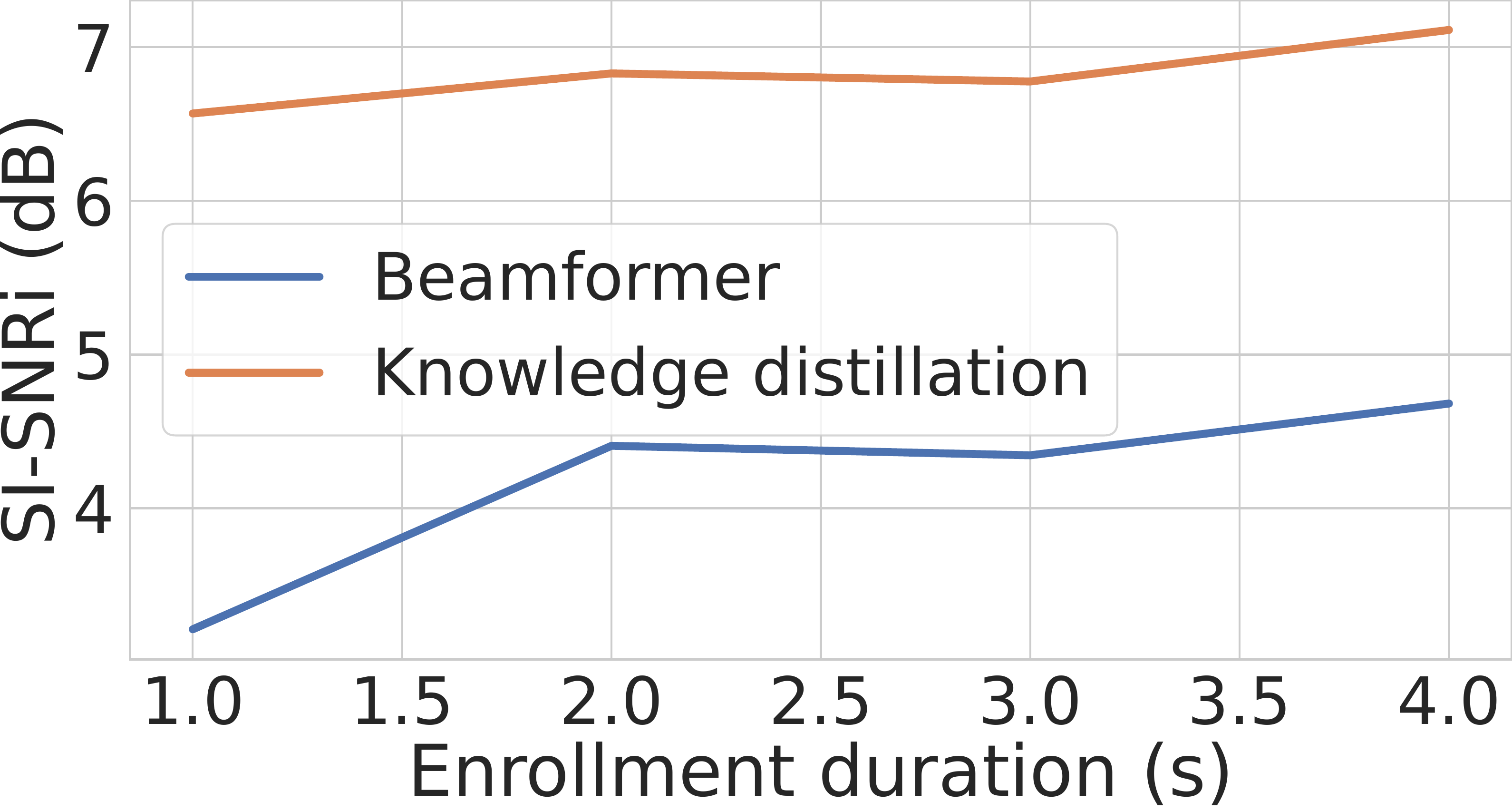}
        \caption{}
        \label{duration_kd}
    \end{subfigure}

    \begin{subfigure}[t]{0.45\textwidth}
        \centering
        \includegraphics[height=1.1in]{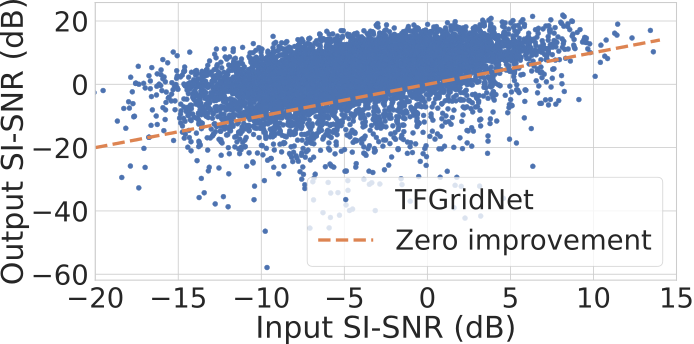}
        \caption{}
        \label{scatter}
    \end{subfigure}
    ~
    \begin{subfigure}[t]{0.45\textwidth}
        \centering
        \includegraphics[height=1.1in]{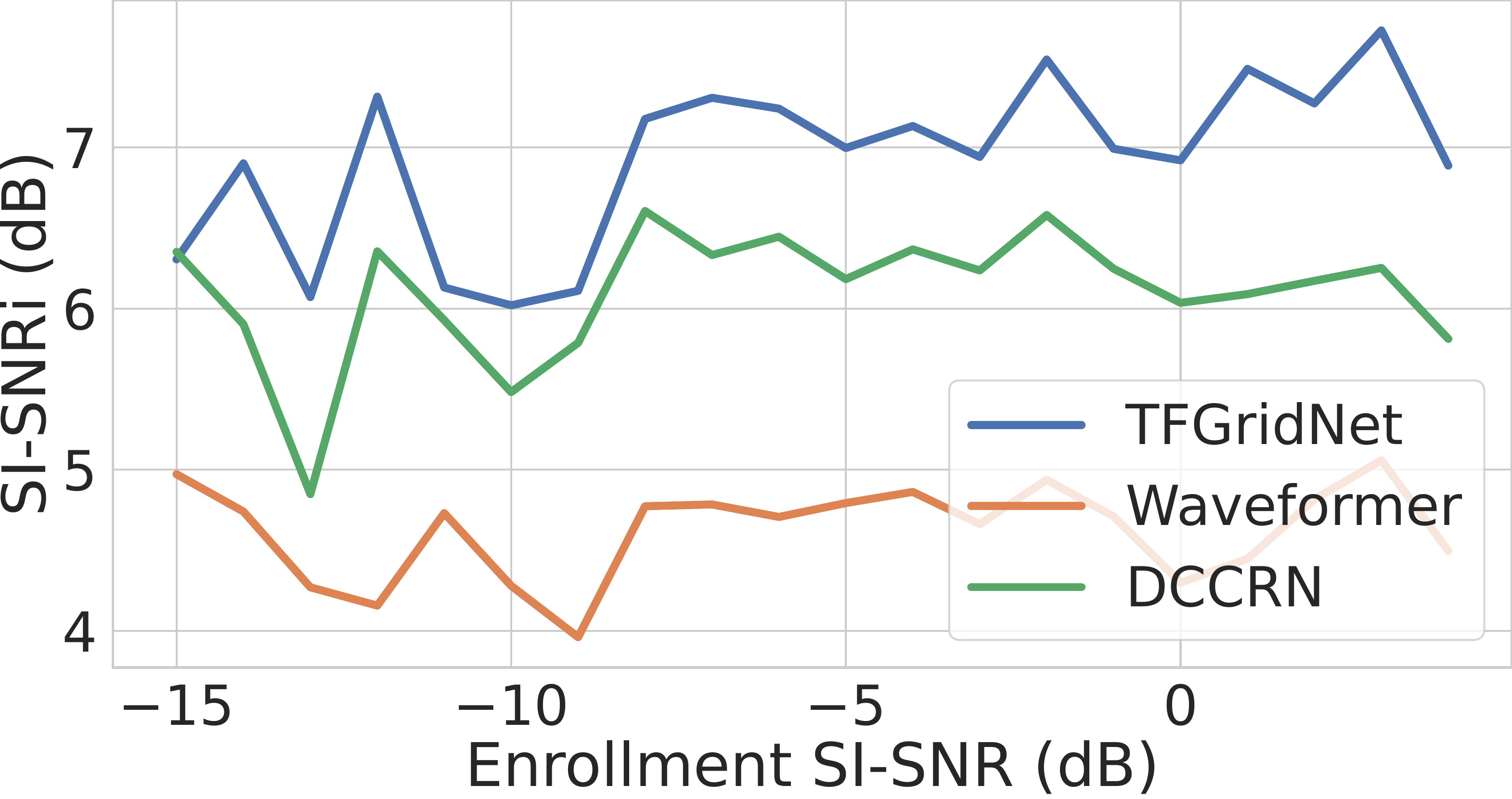}
        \caption{}
        \label{enroll_v_sisnr}
    \end{subfigure}
    \vskip -0.16in
    \caption{(a) Effect of similarity between target and interfering speakers. (b) The gap between beamformer based noisy enrollments and knowledge distillation is larger with shorter enrollments. (c) Model shows improvement across all levels of input noise. {(d) Model is robust to noise  in the enrollment since both enrollment methods are trained to predict speaker embedding in the direction the wearer is looking at,  while ignoring sources in other directions.}}
    \vskip -0.15in
\label{quant_analysis}
\end{figure*}

{In addition, since the original TFGridNet is not designed for streaming applications,  we re-implemented it as a cached, streaming model without any other optimizations and compare our final model's inference time with this streaming TFGridNet implementation running on PyTorch and ONNX Runtime. Table.~\ref{runtime_comparison} shows that  executing inference on the Orange Pi is significantly faster if we use ONNX Runtime, and when we further optimize the model as described in ~\xref{sec:realtime}, we can further achieve a 41.1\% reduction in the inference time with only a 1.5\% reduction in the SI-SNRi.}

In Fig. \ref{intefer_speech}, we measure the performance of the extracted target speech  quality at different interfering speech similarities. Interfering speech similarity is measured as the cosine distance between embeddings computed on the target speech and interfering speech. As observed in the plot, all architectures suffer a near-linear drop in performance when high-level speech characteristics of interfering speech are similar to that of the target speech.
  We observe a direct correlation between the performance and the embedding similarity -- cosine similarity between embeddings estimated on noisy enrollments and embedding estimated on clean enrollments. Compared to clean enrollments, the beamformer  method for noisy enrollments resulted in a 2.4-2.9 dB performance drop across all architectures. The knowledge distillation method  resulted in only a 0.3-0.6 dB drop in performance relative to the clean enrollment.

 To understand this disparity, in Fig.~\ref{duration_kd} {we plot the SI-SNRi as a function of the duration of the target speaker utterance in the noisy enrollment signal. We use 5s-long noisy enrollment signals throughout the experiment but only vary the length of the target speaker's component in the noisy enrollment.} We observe a higher gap with shorter enrollment durations compared to longer enrollments. This is due to the fact that when the target speaker is not present in the enrollment signal for a significant fraction of the time, the faint signal of interferer speech might result in the embedding getting closer to the interferer speaker. The knowledge distillation model appears to be more robust since it is trained to predict only the target embeddings even in such scenarios. While these are results aggregated over the entire test set, we also observed that the model results in a positive improvement in the target speaker's signal quality for close to 90\% of the samples. In Fig.~\ref{scatter} we plot the output SI-SNRi of all samples as a function of input SI-SNR. All the samples that are above the $x=y$ line resulted in a positive improvement. {Further, in Fig.~\ref{enroll_v_sisnr}, we quantify the effect of noisy enrollment signals on the output SI-SNRi. The plot shows that noisy  enrollment signals only have a small effect on the output SI-SNRi across a range of input enrollment SI-SNRs.}

 Finally, in Fig. \ref{ang_vel} and \ref{ang_err}, we measure the performance of a fine-tuned model for angular velocities ranging from 10 deg/s to 80 deg/s and errors in enrollment signal ranging from 5  to {18} degrees. We compare the performance of the fine-tuned system with the version that is not fine-tuned. {We can observe that the fine-tuned system not only performs better}, but also is more robust in the presence of motion and enrollment angular error.


\section{Limitations and Discussion}
While our focus in this paper has been on extracting a target speaker and playing it back into the hearables, one could  also train the  system to {\it remove} a target speaker from a mixture of sounds. This can be helpful in scenarios where say  you want to filter out one person's disruptive speech while still hearing everyone else.  

In our target applications, the user is interested in listening to a single speaker in a crowded environment. This is by itself a  common scenario across multiple use cases. Future work could extend this to support multiple target speakers by enrolling each of them and retraining our network to use multiple speaker enrollments. {One approach is to run multiple instances of our network  for multiple speakers. This, however, would come at a significant on-device compute cost. Instead, training a single  network to extract multiple speakers given some aggregated multi-speaker embedding may produce a system that more efficiently handles multiple speakers.}

An assumption we make  is that during the short enrollment phase, there are no other {\it strong}  interfering  speakers in the same direction as the target speaker. We note two key points: 1) If the target speaker is mobile during the enrollment phase, which our design supports, it reduces the probability of having another strong interfering speaker in the same direction, for the whole enrollment duration. 2) Even in static scenarios, we can train the network to focus on either the closest and/or loudest speaker in the direction, the wearer is looking at, during enrollment. Exploring this would be an interesting avenue for future research.

\begin{figure}[t!]
    \centering
    \begin{subfigure}[t]{0.99\columnwidth}
        \centering
        \includegraphics[width=0.8\columnwidth]{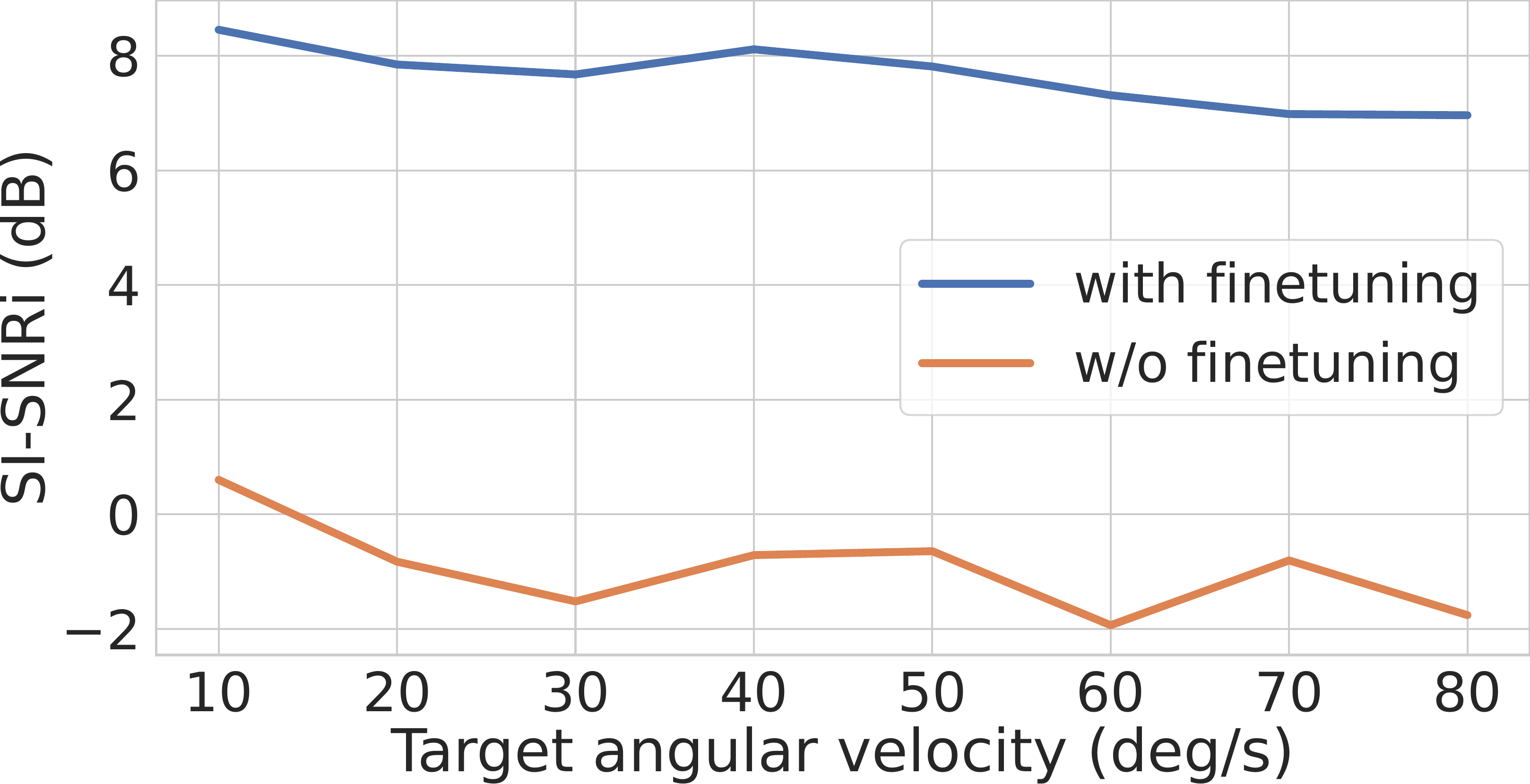}
        \caption{}
        \label{ang_vel}
    \end{subfigure}
    \begin{subfigure}[t]{0.99\columnwidth}
        \centering
        \includegraphics[width=0.8\columnwidth]{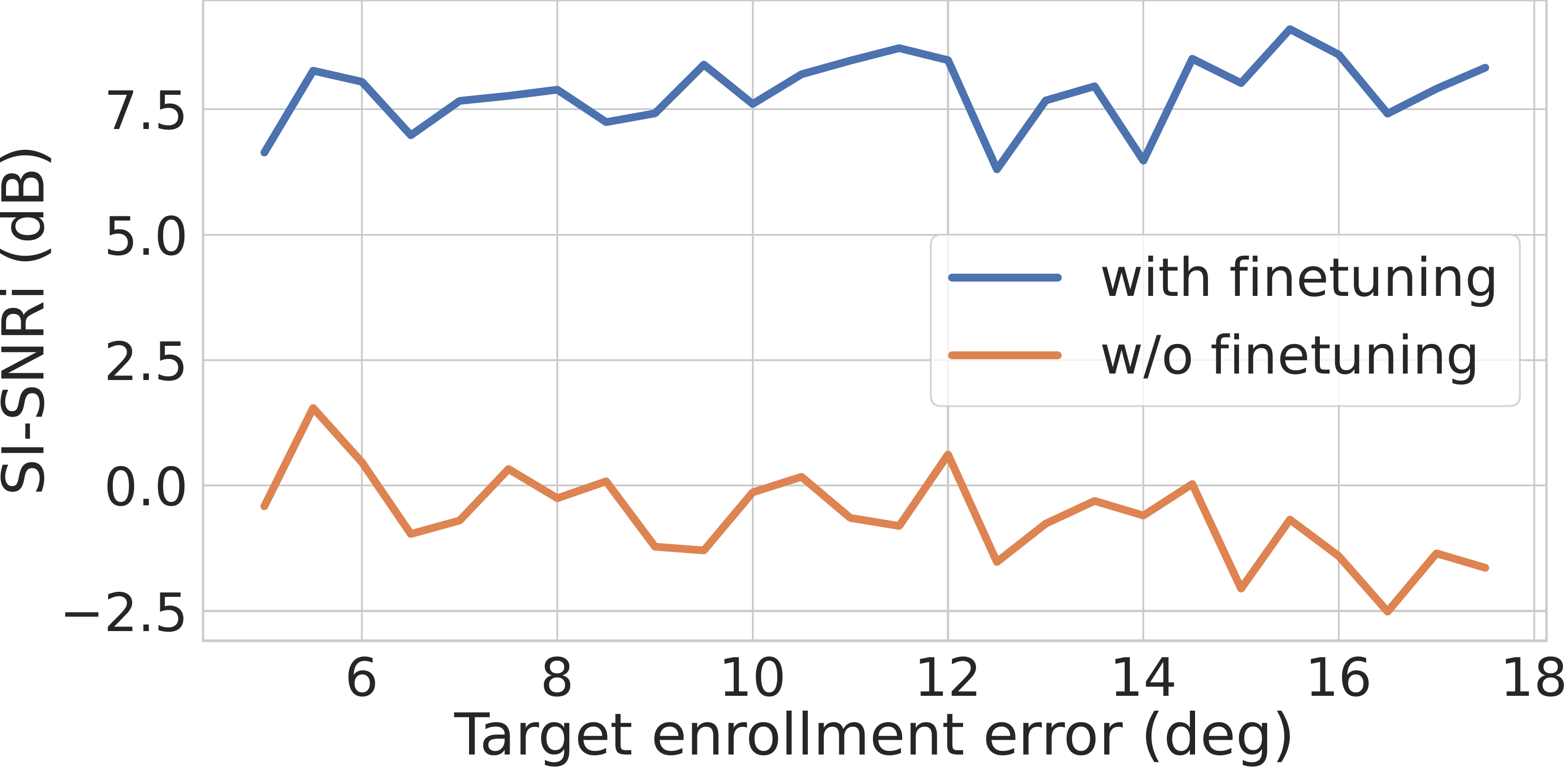}
        \caption{}
        \label{ang_err}
    \end{subfigure}
    \vskip -0.15in
    \caption{Comparison with and without fine-tuning, when relative motion and {enrollment angle error} is present.}
    \vskip -0.15in
\label{ft_analysis}
\end{figure}

Unlike our target speech-hearing network, our enrollment network is  not designed to run on-device. However, it can generate the speaker embedding in hundreds of milliseconds  on a cloud GPU. Having the enrollment network run on a more powerful GPU  (e.g.,  cloud or edge computing server) is an acceptable  choice  since it is a one-time process. 

The speech characteristics  of humans may change with factors such as aging, change in health status and emotions~\cite{tse-overview}, which is an issue with the general problem of target speaker extraction. We however note that in our use case, the wearer uses the binaural hearable to capture an enrollment example of the target speech right before it is being used to extract the target speaker. As a result some of these factors will likely not change in this short  duration.

As illustrated in Fig.~\ref{intefer_speech}, the greater the similarity between the target speaker and one of the interfering speakers, the more  challenging it becomes to   completely eliminate the interfering speaker.  One way to enhance the system's robustness involves using multiple enrollment signals at different time points instead of a continuous one. For instance, if users perceive unsatisfactory signal quality, they can use the enrollment interface to gather additional binaural data, which the model can potentially use to more uniquely identify the target speaker. {Another approach entails leveraging embedding models trained on an even larger scale speech datasets such as LibriLight \cite{librilight} that could be capable of achieving close-to-human-level speaker recognition to in detecting subtle differences between similar-sounding speakers.} Note that  the embedding model does not need on-device execution and only runs occasionally. 


We demonstrated target speech hearing using off-the-shelf binaural noise-canceling headsets. Active noise-canceling (ANC) headsets use an external microphone to capture environmental noise. The headsets then generate an anti-noise signal that cancels out the external sounds while using an internal microphone as feedback to generate the anti-noise signal. However, since users can play music and take phone calls on these headsets, these ANC systems also take the digital audio being played by the headset speakers as input to ensure that the digital audio does not get canceled. Since we play the target speech through the headset speakers, the headset system is already designed not to cancel it, and hence, we were able to demonstrate the feasibility of our design in~\xref{sec:noisecanceling}. It would be interesting to explore whether future ANC headsets can be  designed to account for target speech hearing to further improve performance.

Finally, we prototyped our system using off-the-shelf binaural headsets connected to an embedded IoT CPU. We note that the IoT CPU platform supports specialized neural processing units (NPUs) that we are not currently using in our implementation. However, these NPUs can potentially further improve the processing latency of our neural networks. Furthermore, recent advances in neural accelerators suggest that commercial devices designed for target speech hearing may likely be incorporated on these accelerators  to minimize power consumption and latency.

\section{Conclusion}

This paper makes an important   contribution to the overarching vision of intelligent hearables, where future headsets and earbuds can augment the human auditory perception with  artificial  intelligence, enabling users to actively manipulate their acoustic surroundings in real-time, giving them the ability to select and hear   sounds based on user-defined characteristics, such as speech traits.


Towards this vision, we introduce the concept of target speech hearing using noisy examples on hearables that allows a user to focus on a specific speaker, given their speech characteristics,  while reducing interference from other speakers and noise.  We make three key technical contributions to achieve this new capability for hearables: 1) an enrollment interface that uses a noisy, binaural recording of the target speaker to generate a speaker embedding that captures that traits of the target speaker, 2) a real-time neural network that runs on embedded IoT CPU to extract the target speaker  given the speaker embedding, and 3)  a training methodology that uses synthetic data and yet allows our system to generalize to real-world unseen speakers, indoor and outdoor environments as well as support mobility. Our in-the-wild evaluations show generalization to real-world unseen indoor and outdoor environments. 

\begin{acks}
The University of Washington researchers are partly supported by the Moore Inventor Fellow award \#10617, Thomas J. Cable Endowed Professorship, and a UW CoMotion innovation gap fund. {This work was facilitated through the use of computational, storage, and networking infrastructure provided by the HYAK Consortium at the University of Washington.}
\end{acks}

\bibliographystyle{ACM-Reference-Format}
\bibliography{ref}

\end{document}